\documentclass[11pt,a4paper]{article}

\pdfoutput=1

\usepackage{bm}

\usepackage{graphicx}

\usepackage{multirow}

\usepackage{cite}

\usepackage{hyperref}

\usepackage{geometry}
\usepackage{appendix}

\usepackage{authblk}

\title{Anisotropic universe with anisotropic sources}

\date{ }

\author[]{Pavan K. Aluri\thanks{aluri@iucaa.ernet.in (Now at IUCAA, Pune - 411007, India)}}
\author[]{Sukanta Panda\thanks{sukanta@iiserb.ac.in}}
\author[]{Manabendra Sharma\thanks{manabendra@iiserb.ac.in}}
\author[]{Snigdha Thakur\thanks{snigdha@iiserb.ac.in}}
\affil[]{Department of Physics, IISER Bhopal, Bhopal - 462023, India}

\begin{document}

\maketitle

\begin{abstract}
  We analyze the state space of a Bianchi-I universe with anisotropic
  sources. Here we consider an extended state space which includes null
  geodesics in this background. The evolution equations for all the state
  observables are derived. Dynamical systems approach is used to study the
  evolution of these equations. The asymptotic stable fixed points for all
  the evolution equations are found. We also check our analytic results
  with numerical analysis of these dynamical equations. The evolution of
  the state observables are studied both in cosmic time and using a
  dimensionless time variable. Then we repeat the same analysis with a
  more realistic scenario, adding the isotropic (dust like dark) matter and
  a cosmological constant (dark energy) to our anisotropic
  sources, to study their co-evolution. The universe now approaches a
  de~Sitter space asymptotically dominated by the cosmological constant.
  The cosmic microwave background anisotropy maps due to shear are also
  generated in this scenario, assuming that the universe contains
  anisotropic matter along with the usual (dark) matter and vacuum (dark) energy
  since decoupling. We find that they contribute dominantly to the CMB quadrupole.
  We also constrain the current level of anisotropy and also search for
  any cosmic preferred axis present in the data. We use the Union~2
  Supernovae data to this extent. An anisotropy axis close to the
  mirror symmetry axis seen in the cosmic microwave background data from
  Planck probe is found.
\end{abstract}

\section{Introduction}
In standard cosmology it is assumed that the space-time is homogeneous
and isotropic. After the discovery of temperature anisotropies of the
Cosmic Microwave Background (CMB) and accelerating expansion of the
universe, a standard cosmological model describing universe dominated
by cold dark matter (CDM) and dark energy in the form of a cosmological
constant ($\Lambda$) is formulated, known as $\Lambda$CDM model or cosmic
concordance model. Fluctuations in the temperature of CMB radiation are
statistically isotropic in this model \cite{cobewmap}. A central assumption
in our standard model of cosmology (and in most cosmological models) is that
the universe is homogeneous and isotropic upto small perturbations,
and thus described by a perturbed Friedmann-Robertson-Walker (FRW)
metric. Observations show that the temperature of the CMB is isotropic
to a remarkable degree indicating that our universe is close to a
Friedmann-Lemaitre (FL) model. According to Ehlers, Geren and Sachs (EGS)
theorem \cite{EGS1968}, if the CMB temperature were exactly isotropic
about every point in space-time, then the universe has to be
exactly an FL model. This result is not directly applicable to CMB,
because, CMB radiation is not exactly isotropic. The presence of
temperature anisotropies in the CMB from the observations made by
COBE/WMAP satellites are inconsistent with an exact homogeneous and
isotropic FRW model \cite{SME1995}.

The current CMB data supports an inflationary Big Bang model of
cosmic origin for our universe. However at large angular scales,
there are some anomalies observed in CMB data, such as, a low value
for quadrupole power, alignment of quadrupole and octopole modes
roughly in the direction of Virgo cluster, ecliptic north-south
power asymmetry, parity asymmetry between even and odd multipoles,
almost zero correlations on large angular scales of CMB and an
anomalous cold spot ($\sim 10^\circ$ diameter) in the southern
galactic hemisphere of the CMB sky \cite{cmbanomalies}. Several
solutions to these CMB anomalies have been put forth
such as foregrounds/systematics, anisotropic space-times and exotic
topologies \cite{anomalysolutions}. Some of these supposed deviations
seen in CMB data were addressed by the WMAP and Planck science teams
also \cite{WMAP7anomalies,PlanckAniso}. These deviations may be an
evidence that we live in a globally anisotropic universe.

Generally, in a global anisotropic universe, during
inflation, shear decreases and eventually it goes over to an isotropic
phase with negligible shear \cite{anisotropicinflation}. In order
to produce any substantial amount of shear in recent times one needs
to induce anisotropy in the space-time. One way to induce anisotropy
is to have anisotropic matter present at the last scattering surface.
It was shown earlier that the power suppression in CMB quadrupole,
without affecting higher multipoles, can be explained by assuming
anisotropic matter (magnetic fields) \cite{Campanelli0607}. Earlier
works on anisotropic cosmological models with anisotropic stresses
can be found in Ref.~\cite{BarrowWorks}.
Anisotropic dark energy as a possible solution to the cosmic
acceleration as well as the large scale CMB anomalies is studied 
in Ref.~\cite{AnisoDE}. The anisotropic sources can
be a uniform magnetic field, cosmic strings or domain walls.
A nanogauss scale magnetic field could be present today which would
have been produced during inflation due to a Lorentz-violating term
to the photon sector \cite{Campanelli2009}. We call this as
``Lorentz Violation generated magnetic field (LVMF)''. Other
possibilities such as a Maxwell type vector field coupled to a scalar
field has been studied Ref.~\cite{Thorsud2012}.

In this paper our approach is not to deal with metric approach directly,
rather we adhere to an approach similar to orthonormal frame
formalism \cite{Wainwright1997}. The paper is organized as follows.
In section~\ref{sec:BianchiI}, we write down the Einstein's equations
in cosmic time. Null geodesic equations in cosmic time are studied in
section~\ref{sec:geodesic}. In section~\ref{sec:fixedpoint}, fixed
point analysis of all the evolution equations is carried out in terms of a dimensionless
time variable ($\tau$), to study the asymptotic evolution of the state
observables. A more realistic model including the ordinary dust like (dark) matter
and a cosmological constant (dark energy) in addition to the 
anisotropic sources is considered in section~\ref{sec:realscenario}.
Then, in section~\ref{sec:cmbpattern}, we show the temperature
patterns for CMB due to co-evolution of individual anisotropic sources
along with isotropic matter and vacuum energy.
In section~\ref{sec:snconstr}, we obtain constraints on the
fractional energy densities of various
components of the realistic model described earlier, the level of
shear, and also determine a cosmic preferred axis if present.
Finally, in section~\ref{sec:conclusion}, we conclude our work.

\section{Bianchi-I universe}\label{sec:BianchiI}
In our approach to study dynamical systems, we formulate the evolution
equations in terms of few dimensionless state observables. It is a
very invaluable tool to obtain qualitative information about the
solutions of the state space of Bianchi universes (see Ref.~\cite{Wainwright1997}
and references therein). Here we concentrate only on Bianchi-I model.
First we derive the Einstein's equations in metric approach. Then we
write them in terms of state observables analogous to the orthonormal
frame formalism \cite{EllisMacCallum1969}. In this approach one
writes the field equations as first order differential equations.
An advantage of this orthonormal frame approach is that the
derivation of geodesic equation is easy compared to the metric
approach.

We start with a Bianchi-I line element with a residual planar symmetry
in the $yz-$plane as
\begin{equation}
ds^2 = dt^2  - a(t)^2 dx^2 - b(t)^2 (dy^2 + dz^2).
\label{eq:metric}
\end{equation}
We choose the diagonal energy momentum tensor of the form
$T_{\nu}^{\mu}=(\rho, -p_a, -p_b, -p_b ),$ where $p_a=w_a \rho$
and $p_b =w_b \rho$. The four kinds of anisotropic matter we study
here are given in Table~[\ref{tabl:matter}]. The magnetic field
configuration considered here as an anisotropic source is well
known \cite{BarrowWorks}. We, then consider anisotropic matter configurations
due to topological defects such as cosmic strings and domain walls.
The energy-momentum tensors corresponding to these sources
are given in appendix~\ref{apdx:A}.
Finally, a magnetic field generated due to a Lorentz-violating term in the
photon sector, abbreviated as LVMF, is also studied as an
anisotropic source \cite{Campanelli2009}.
Here we assume that all these sources are of primordial origin and
were produced through breaking of discrete symmetries during phase transitions
in the early universe. A distribution of cosmic strings
and domain walls may quickly grow anisotropic. We analyse them towards
our motivation to study the anisotropic sources and their (asymptotic)
evolution, and co-evolution with the (dust like) dark matter and
dark energy ($\Lambda$), relevant at late times. Eventually we also obtain
constraints on these anisotropic sources using Type Ia supernova data.

\begin{table}
\centering
\begin{tabular}{|l|l|l|r|}
  \hline
  Matter            & $w_a$ & $w_b$ \\
  \hline
  Cosmic String     & -1    & 0     \\
  Domain Walls      &  0    & -1    \\
  LVMF              &  1    & 0     \\
  Magnetic Field    & -1    & 1     \\
  \hline
\end{tabular}
\caption{Equation of state parameters for different anisotropic sources considered
         in this paper.}
\label{tabl:matter}
\end{table}

The Einstein's equations for the metric given in Eq.~[\ref{eq:metric}] are
\begin{eqnarray}
 \label{eq:einstein1}
 2 H_a H_b + H_b^2 &=& 8 \pi G \rho \,, \nonumber \\
 2 \dot{H_b} + 3 H_b^2 &=& -8 \pi G p_a \,, \\
 \dot{H_a}+ \dot{H_b} + H_a^2 + H_a H_b + H_b^2 &=& -8 \pi G p_b \,, \nonumber
\end{eqnarray}
where $H_a=\frac{1}{a}\frac{da}{dt}$ and $H_b=\frac{1}{b}\frac{db}{dt}$.
The equation of continuity is given by,
\begin{equation}
\dot{\rho} + (H_a +2 H_b) \rho + H_a p_a + 2 H_b p_b =0\,.
\label{eq:rho}  
\end{equation}

Here onwards we choose to set $8 \pi G = 1$ for the rest of this
paper. This can otherwise be seen as rescaling, all the  quantities
in Eq.~[\ref{eq:einstein1}] with $1/\sqrt{8 \pi G}$. Now, we rewrite
the above set of equations, Eq.~[\ref{eq:einstein1}] and [\ref{eq:rho}],
in terms of the average expansion $H =(H_a + 2 H_b)/3$ and shear
$h =(H_b-H_a)/\sqrt{3}$. The Einstein's equations and the equation
of continuity in cosmic time become
\begin{eqnarray}
 \label{eq:einstein2}
 \frac{d H}{dt} &=& - H^2 -\frac{2}{3} h^2 - \frac{1}{6} (\rho + p_a + 2 p_b) \,, \nonumber \\
 \frac{d h}{dt} &=& - 3 H h + \frac{1}{\sqrt{3}} (p_b -p_a) \,, \\
 \frac{d \rho}{dt} &=& -3 H (\rho + \frac{p_a + 2 p_b}{3}) - \frac{2 h}{\sqrt{3}}(p_b-p_a) \,, \nonumber
\end{eqnarray}
and a constraint equation given by
\begin{equation}
H^2 = \frac{\rho}{3} + \frac{h^2}{3} \,.
\end{equation}

\begin{figure}
  \centering
  $
  \begin{array}{c c}
    \includegraphics[width=0.43\textwidth]{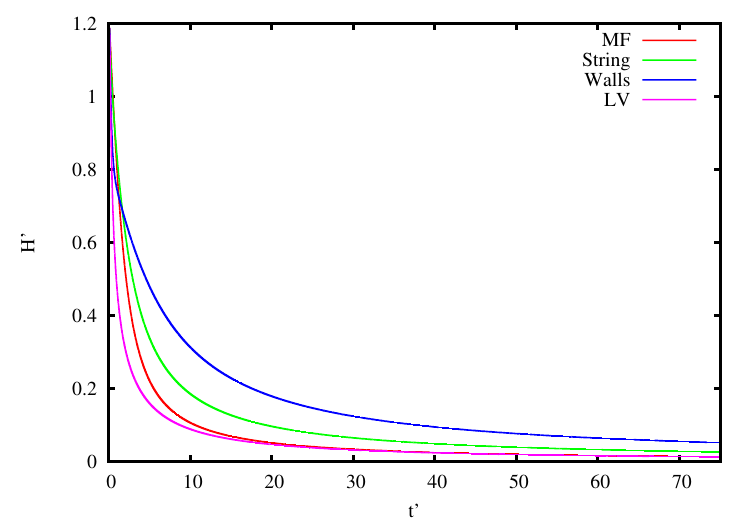} &
    \includegraphics[width=0.43\textwidth]{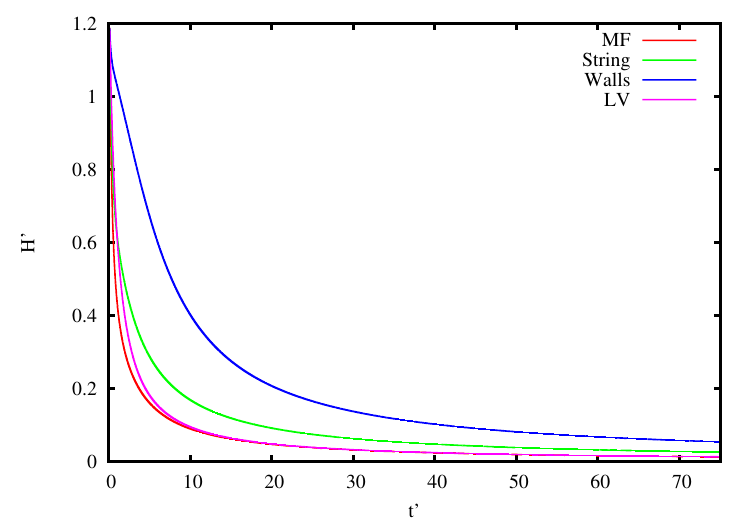}
  \end{array}
  $
  \caption{$H'$ vs. $t'$. In the \emph{left} figure, $h'(0) = 173, h_0/H_0 = 0.01,
           \Omega_0 = 0.1, \rho'(0)= 10$, and in the \emph{right} figure $h'(0) = -173,
           h_0/H_0 = 0.01, \Omega_0 = 0.1, \rho'(0)= 10$.}
  \label{fig:hubble}
\end{figure}

\begin{figure}
  \centering
  $
  \begin{array}{c c}
    \includegraphics[width=0.43\textwidth]{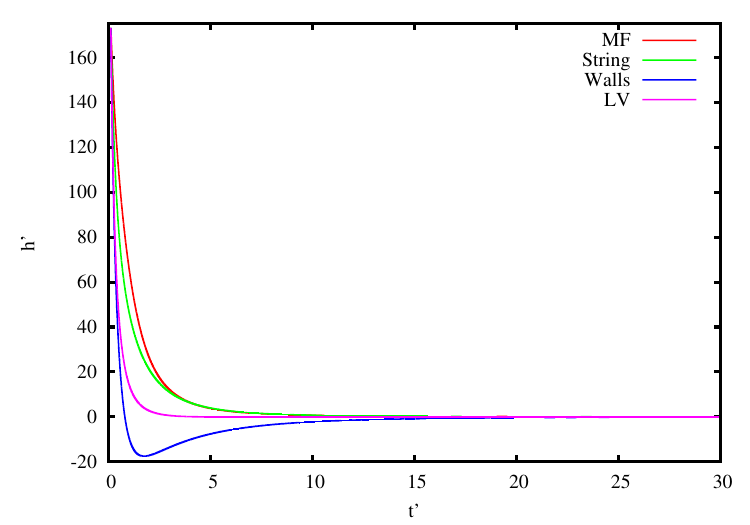} &
    \includegraphics[width=0.43\textwidth]{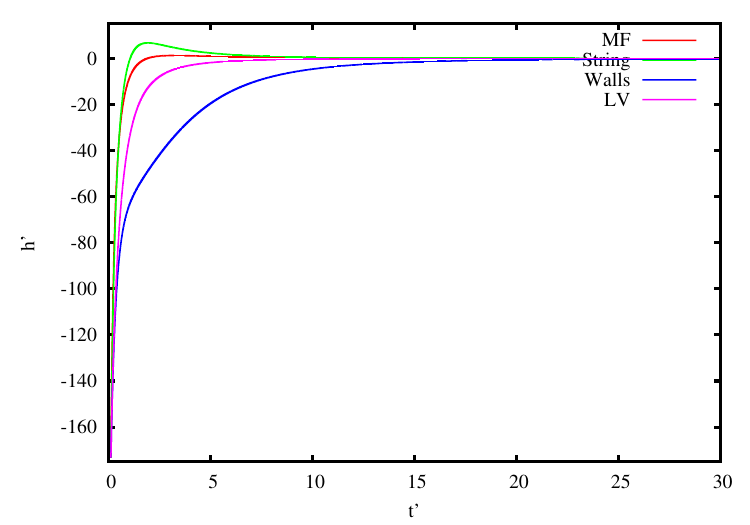}
  \end{array}
  $
  \caption{$h'$ vs. $t'$. In the \emph{left}(\emph{right}) figure, the initial data is same as in
           \emph{left}(\emph{right}) figures of Fig.~[\ref{fig:hubble}].}
  \label{fig:shear}
\end{figure}

\begin{figure}
  \centering
  $
  \begin{array}{c c}
    \includegraphics[width=0.43\textwidth]{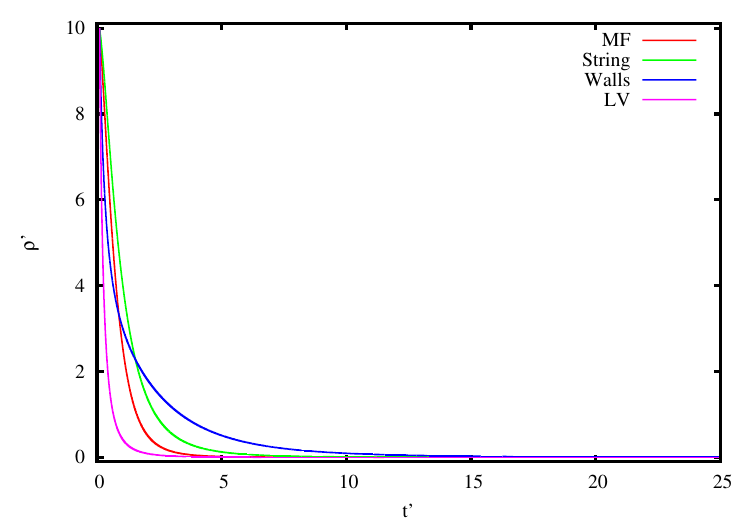} &
    \includegraphics[width=0.43\textwidth]{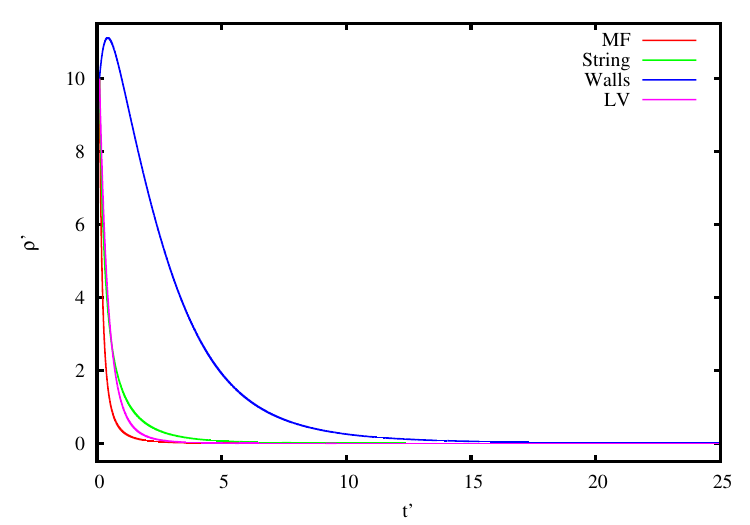}
  \end{array}
  $
  \caption{$\rho'$ vs. $t'$. In the \emph{left}(\emph{right}) figure, the initial data is same as in
           \emph{left}(\emph{right}) figures of Fig.~[\ref{fig:hubble}].}
  \label{fig:rho}
\end{figure}

We now introduce dimensionless variables $H'$, $h'$, $t'$ defined
as
\begin{eqnarray}
 \label{eq:einstein3}
 H'&=& \frac{H}{H_0} \,, \nonumber \\
 h'&=& \frac{h}{h_0} \,, \\
 t' &=& t H_0 \,, \nonumber \\
 \rho' &=& \frac{\rho}{\rho_0} \,, \nonumber
\end{eqnarray} 
where $H_0$, $h_0$ and $\rho_0$ are the current values of Hubble, shear
and anisotropic matter density parameters, respectively. Then, the
equations for the state observables in terms of these dimensionless
variables can be written as
\begin{eqnarray}
 \label{eq:einstein4}
 \dot{H'} &=& - H'^2 - \frac{2}{3} \left(\frac{h_0}{H_0}\right)^2 h'^2 - \frac{1}{2} (1 + w_a + 2 w_b) \Omega_0 \rho' \,, \nonumber \\
 \dot{h'} &=& - 3 H' h' + \sqrt{3} (w_b - w_a) \frac{H_0}{h_0} \Omega_0 \rho' \,, \\
 \dot{\rho'} &=& -3(1 + \frac{w_a + 2 w_b}{3}) H' \rho' - \frac{2(w_b-w_a)}{\sqrt{3}} \frac{h_0}{H_0}h'\rho' \,, \nonumber
\end{eqnarray}
where an overdot represents differentiation with respect to $t'$
and $\Omega_0={\rho_0}/{3 H_0^2}$. Now, we solve the above equations
numerically for all the cases listed in Table~[\ref{tabl:matter}].
The evolution of the dynamic variables $H'$, $h'$ and $\rho'$ are
plotted in Fig.~[\ref{fig:hubble}], [\ref{fig:shear}] and
[\ref{fig:rho}]. For all these anisotropic matter sources, it turns
out that $H'$ decreases with increasing time $t'$. Uniform magnetic
field and LVMF matter have almost similar rate of expansion, where
as walls and strings have a higher rate of expansion.
From Fig.~[\ref{fig:shear}], shear becomes negligible at late times
and it decreases slowly for walls compared to the other three sources.
Isotropy appears to set in at late times for all cases of
anisotropic matter. A fixed point analysis will shed some
light on the actual evolution. Detailed inspection of attaining 
isotropy at late times will be done in section~\ref{sec:fixedpoint}.
We observe from Fig.~[\ref{fig:rho}] that, the energy density also
decreases much faster for magnetic field, LVMF and strings compared
to walls.

From Fig.~[\ref{fig:shear}], it can be observed that evolution
of shear depends on the initial conditions. Signature change of shear
can be seen for certain kind of anisotropic matter. This signature
change may have implications in early universe. From the left plot
of Fig.~[\ref{fig:shear}], we see that this change occurs for walls, 
whereas the right plot of Fig.~[\ref{fig:shear}] shows a similar signature
change but for cosmic strings, depending on a positive/negative shear
in the beginning. However late time behaviour of $H'$, $h'$ and $\rho'$ in
cosmic time are independent of initial conditions for all these sources.

\section{Null Geodesic evolution}\label{sec:geodesic}
As we know the geodesic equations are second order differential
equations in metric approach. If we represent the same equation in
terms of $H$ and $h$, then the geodesics can be written as first
order differential equations given by \cite{Nilsson1999}
\begin{eqnarray}
 \label{eq:geodesic1}
 \frac{d \epsilon}{dt} &=& - H \epsilon -\frac{(\epsilon^2 - 3 k_1^2)h}{\sqrt{3} \epsilon} \,, \nonumber \\
 \frac{d k_1}{dt} &=&  \left(-H + \frac{2h}{\sqrt{3}}\right) k_1 \,, \nonumber \\
 \frac{d k_2}{dt} &=&  \left(-H -\frac{h}{\sqrt{3}}\right) k_2 \,, \\
 \frac{d k_3}{dt} &=&  \left(-H - \frac{h}{\sqrt{3}}\right) k_3 \,, \nonumber
\end{eqnarray}
where $\epsilon=k_0$ is the energy of photon, and $k_1$, $k_2$ and $k_3$  are
the three components of photon momentum ($\vec{k}$). The variables $\epsilon$ and
$k_i$ ($i$=1,2,3) satisfies the constraint equation
\begin{equation}
\epsilon^2 = |\vec{k}|^2 \,.
\label{eq:energy1}
\end{equation}
In terms of cosmic time $t'$ the above equations become
\begin{eqnarray}
 \label{eq:geodesic2}
 \dot{\epsilon} &=& - H' \epsilon - \frac{h_0}{H_0} \frac{(\epsilon^2 - 3 k_1^2)h'}{\sqrt{3}\epsilon} \,, \nonumber \\
 \dot{k_1} &=&  \left(-H' + 2 \frac{h_0}{H_0}  \frac{h'}{\sqrt{3}}\right) k_1 \,, \nonumber \\
 \dot{k_2} &=&  \left(-H' - \frac{h_0}{H_0} \frac{h'}{\sqrt{3}}\right) k_2 \,, \\
 \dot{k_3} &=&  \left(-H' - \frac{h_0}{H_0} \frac{h'}{\sqrt{3}}\right) k_3 \,, \nonumber
\end{eqnarray}
where an overdot here represents differentiation with respect to $t'$.
\begin{figure}
  \centering
  $
  \begin{array}{c c}
    \includegraphics[width=0.43\textwidth]{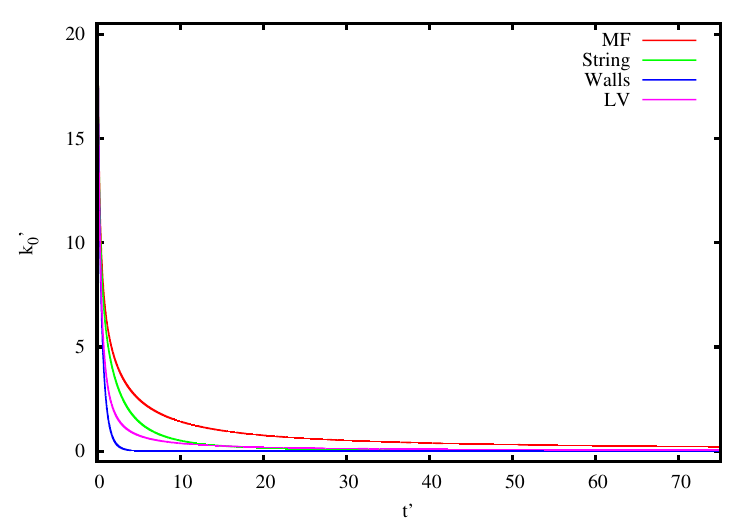} &
    \includegraphics[width=0.43\textwidth]{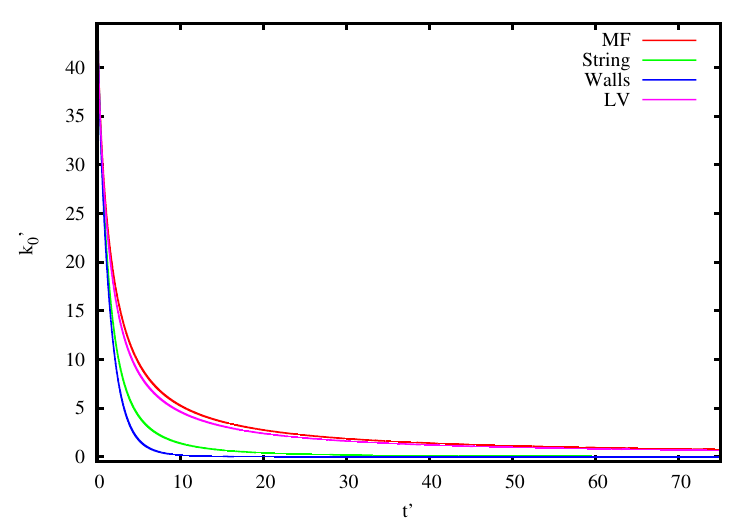}
  \end{array}
  $
  \caption{$k_0$ vs. $t'$. The initial data is same as in Fig.~[\ref{fig:hubble}] with
          $\epsilon(0) = k_0 = 20$ (\emph{left} - Case 1) and, $\epsilon(0) = k_0 = 20$ and
          $k_1(0)=k_3(0)=25.3$ (\emph{right} - Case 2).}
  \label{fig:k0}
\end{figure}
\begin{figure}
  \centering
  $
  \begin{array}{c c}
    \includegraphics[width=0.43\textwidth]{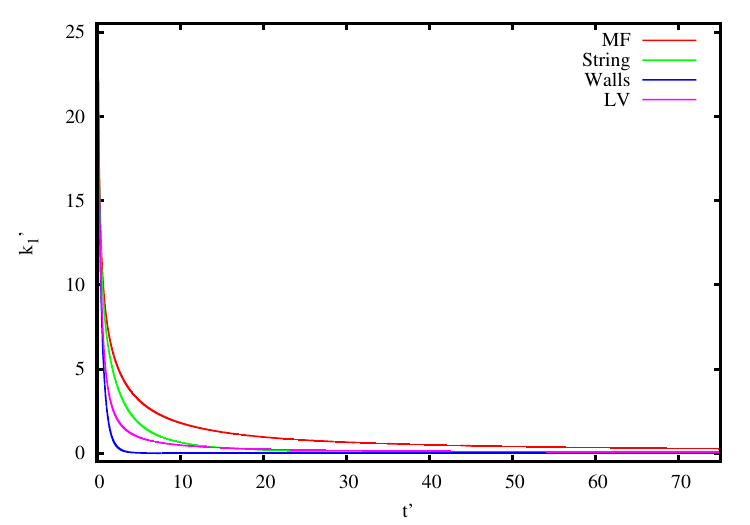} &
    \includegraphics[width=0.43\textwidth]{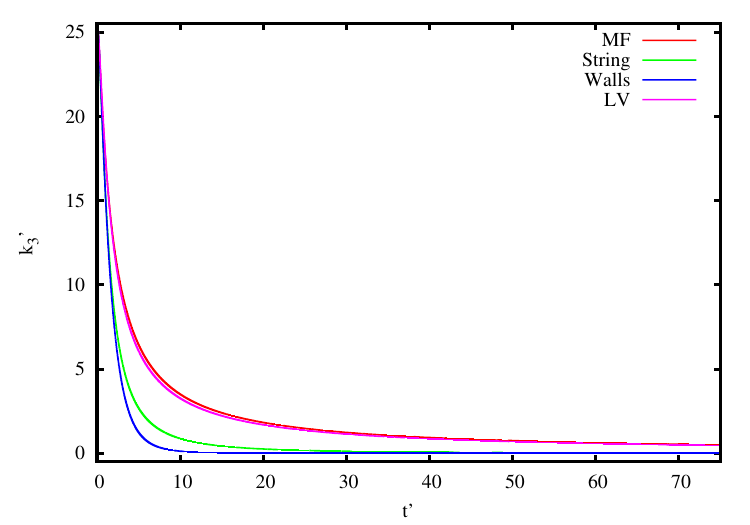}
  \end{array}
  $
  \caption{Shown here are $k_1$ vs. $t'$ (\emph{left}) and $k_3$ vs. $t'$ (\emph{right}).
           The initial data is same as in Fig.~[\ref{fig:hubble}] except for $k_1(0)=k_3(0)=25.3$.
           These plots correspond to Case 2.}
  \label{fig:k1_k3}
\end{figure}

In order to solve the null geodesics, Eq.~[\ref{eq:geodesic2}], one
needs to take care that the solutions simultaneously
satisfy the constraint equation, Eq.~[\ref{eq:energy1}].
We solve these geodesic equations numerically for the following
two cases.\\
{\bf Case 1 :} In this case only $k_1$ is taken to be nonzero, and
$k_2$ and $k_3$ are zero. Initial values are taken as $k_0(0) = 20$ and
$ k_2(0) = k_3(0) = 0 $.
The constraint equation, Eq.~[\ref{eq:energy1}], gives $ k_1 = \pm k_0 = \pm \epsilon$.
Here we choose the positive value for $k_1$. The evolution of $k_0$ is
shown in the left plot of Fig.~[\ref{fig:k0}] for an initial values of $k_0(0) = 20$,
for the four anisotropic sources we are considering.
Here we find that, if the initial value of $k_1$ is positive then for the
entire evolution it remains positive.
It's also true for negative initial values of $k_1$, where it remains negative
for the whole evolution. Other initial values
are same as those of Fig.~[\ref{fig:hubble}], [\ref{fig:shear}] and [\ref{fig:rho}].
We observe that $k_0$ decreases slowly for magnetic field and faster for
walls. This can be understood from Fig.~[\ref{fig:hubble}] where $H'$
decreases slowly for walls and faster for a magnetic field.\\
{\bf Case 2:}  In this case we take $k_1(0) = k_3(0) = 25.3$ and,
$k_2 = k_3$ following the residual planar symmetry of our metric.
Other initial values are same as Case~1.
The evolution of $k_0$ is shown in the right plot of right plot
of Fig.~[\ref{fig:k0}], and $k_1$ and $k_3$ are shown in Fig.~[\ref{fig:k1_k3}].
Here also, we observe that $k_0$ decreases slowly for magnetic field
and faster for walls. $k_1$ and $k_3$ also evolve in a similar manner,
except for the interchange of evolution of cosmic strings and LVMF.

\section{Fixed point analysis of the evolution equations}
\label{sec:fixedpoint}
In the previous section, we studied the evolution of state observables
in terms of a time variable analogous to cosmic time.
In order to study the asymptotic evolution of our Bianchi-I universe
with various anisotropic sources considered here, we resort to
dynamical systems approach. So, we cast all the evolution equations in
terms of dimensionless variables which give us a set of dynamical equations.
The advantage of working in this approach is that we can show the
cosmological dynamics insensitive to the initial conditions. Here we
define the dimensionless variables $\tau$ and $\sigma$ as $d\tau/dt = H$
and $\sigma = {h}/{\sqrt{3}H}$, respectively. These are called expansion
normalized time and shear variables. Thus our Einstein's equations
and the continuity equation are given by
\begin{eqnarray}
\label{eq:sigma}
\frac{dH}{d\tau} &=& -H(1+q) \,, \nonumber \\
\frac{d\sigma}{d \tau} &=& \frac{3}{2}\left[-\sigma(1-\sigma^2)+(\frac{2(w_b-w_a)}{3}+w\,\sigma)\Omega\right]
\end{eqnarray}
and
\begin{equation}
\Omega + \sigma^2 = 1 \,,
\label{eq:omega}
\end{equation}
where $q = 2 \sigma^2 + (1+2 w) \Omega/2$, $w = (w_a + 2 w_b)/3$
and $\Omega = {\rho}/{3 H^2}.$

\begin{figure}
 \centering
 \includegraphics[width=0.84\textwidth]{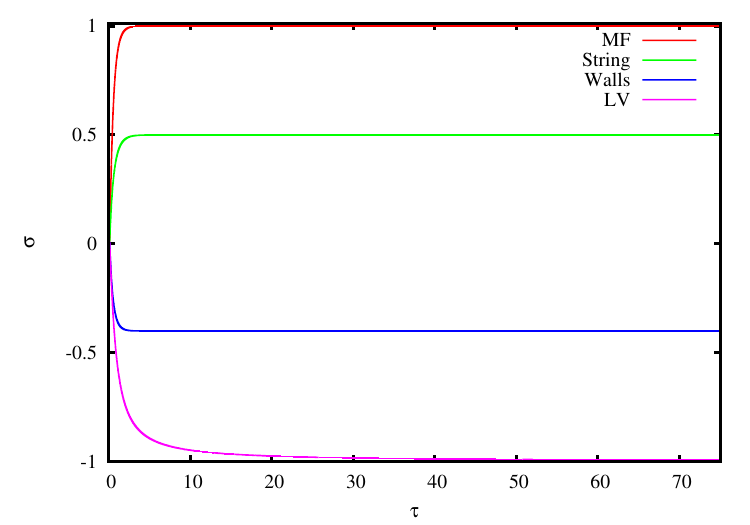}
 \caption{$\sigma$ vs $\tau$ : Late time behaviour of $\sigma$ for
          each of the four anisotropic matter sources considered
          separately, are shown here.}
 \label{fig:sigma}
\end{figure}

\begin{figure}
 \centering
 $
 \begin{array}{c c}
  \includegraphics[width=0.47\textwidth]{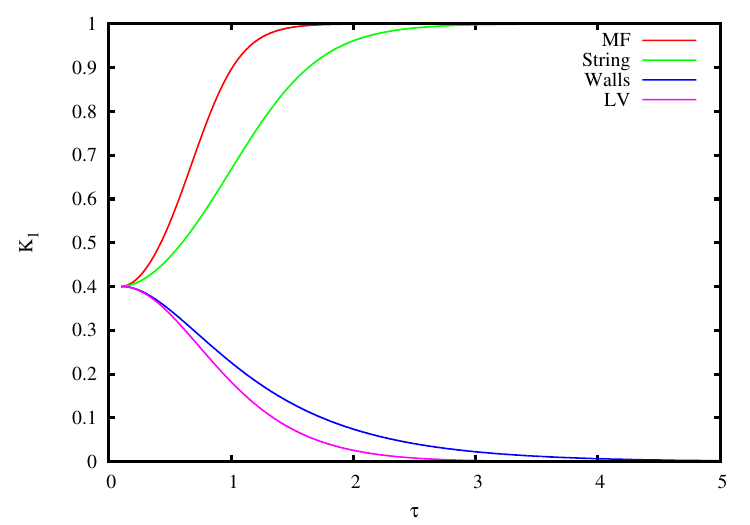} &
  \includegraphics[width=0.47\textwidth]{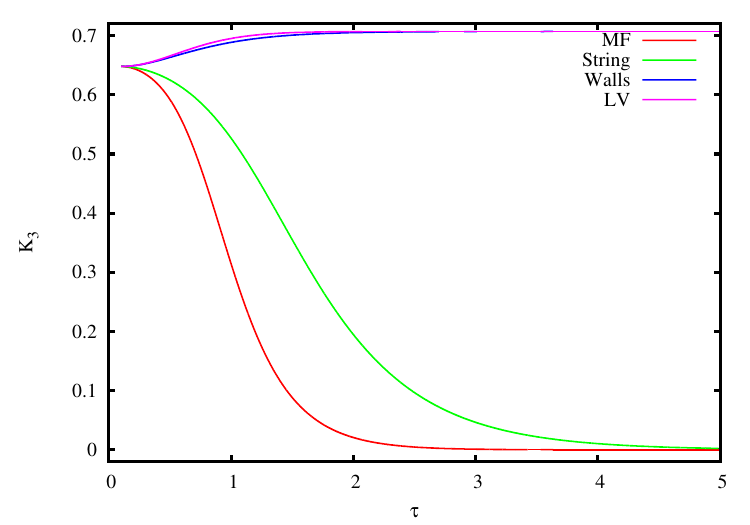}
 \end{array}
 $
 \caption{Evolution of $K_1$ vs. $\tau$ (\emph{left}), and
          $K_3$ vs. $\tau$ (\emph{right}) are shown here
          for the various anisotropic matter sources.
          Here an initial value of $\sigma=0.01$ is used.}
 \label{fig:K1_K3_tau}
\end{figure}

Next, we write the null geodesic equations in terms of dimensionless
variables $K_i = {k_i}/{k_0}$ ($i$=1,2,3). They are given by
\begin{eqnarray}
 \label{eq:geodesic3}
 \frac{d\epsilon}{d\tau} &=& -(1 + s) \epsilon \,, \nonumber \\
 \frac{dK_1}{d\tau} &=&  (s + 2 \sigma) K_1 \,, \nonumber \\
 \frac{dK_2}{d\tau} &=&  (s - \sigma) K_2 \,, \\
 \frac{dK_3}{d\tau} &=&  (s - \sigma) K_3 \,, \nonumber
\label{eq:momentum}
\end{eqnarray}
where $s = (1 - 3 K_1^2)\sigma$. The variables $K_i$ ($i$=1,2,3)
satisfy the constraint equation
\begin{equation}
 |\vec{K}|^2 = 1 \,.
\label{eq:energy2}
\end{equation}
Now, we will analyze the fixed points of the full set of equations
combining Einstein's equations, equation of continuity and the
null geodesic equations \cite{Nilsson1999}. Fixed points for
$\sigma$ can be easily obtained from Eq.~[\ref{eq:sigma}], which
are $\pm 1$ and $2(w_b - w_a)/3(1-w)$. From Eq.~[\ref{eq:geodesic3}],
we have $K_1 = \pm 1$ and $K_2=0$ or $K_1=0$ and $K_2=\pm 1/\sqrt{2}$,
as implied by Eq.~[\ref{eq:energy2}], as fixed points.
All the stable fixed points are
given in Table~[\ref{tabl:fixpt}]. Detailed evaluation of stable
fixed points is given in appendix~\ref{apdx:B}.
Same conclusion can be drawn from solving these equations
numerically. The evolution of $\sigma$ and,
$K_1$ and $K_3$  are plotted in Fig.~[\ref{fig:sigma}] and
[\ref{fig:K1_K3_tau}]. Asymptotically, we can see that $\sigma$,
$K_1$ and $K_3$ evolve towards their corresponding stable
fixed points.
We now clearly see that isotropy can't be attained at late times for any of
the anisotropic sources, considered here, through this fixed point
analysis. This is in contrast to what \emph{appeared} to be the case studying
the physical variables themselves in the preceding sections.
This is one of the advantages of such an analysis. One may not draw
a similar conclusion from our analysis done in cosmic time.
This whole evolution depends on the positive or negative initial values
of $K_1$ and $K_3.$ Once we choose positive initial value for $K_1$ and
$K_3$, then it remains positive for the whole evolution. Similarly it
stays negative for the whole evolution if negative initial values are
chosen.

\begin{table}
 \centering
 \begin{tabular}{|l|l|l|r|}
  \hline
  Matter            & $\sigma_0$ & $(K_1^0, K_2^0, K_3^0)$ \\
  \hline
  Cosmic String     &    1/2   & $(\pm 1, 0, 0)$ \\
  Domain Walls      &   -2/5   & $(0, \pm 1/\sqrt{2}, \pm 1/\sqrt{2})$ \\
  LVMF &   -1     & $(0, \pm 1/\sqrt{2}, \pm 1/\sqrt{2})$ \\
  Magnetic Field    &    1     & $(\pm 1, 0, 0)$ \\
  \hline
 \end{tabular}
 \caption{Stable fixed points of the whole state space of a Binachi~I
          universe for different anisotropic matter sources.}
 \label{tabl:fixpt}
\end{table}

\subsection*{Kasner solutions}
Exact solutions of Kasner type
can be obtained for the Einstein's equations corresponding to the
fixed points of the shear parameter $\sigma$ \cite{Calogero2011}. In order to find
Kasner type solution at stable fixed points we choose scale factors
as $a(t) = A\, t^{\alpha}$ and $b(t)= B\, t^{\beta},$ where A and B are some
positive constants. For these scale factors, we can write the evolution equations as
\begin{eqnarray}
H_a = H(1 - 2 \sigma) \nonumber \\ 
H_b = H(1 + \sigma)
\label{scale}
\end{eqnarray}
and average scale factor $H$ satisfies
\begin{equation}
\frac{dH}{dt} = - (1+q) H^2.
\label{average}
\end{equation}
After substituting the power law form for the scale factors into
Eq.~[\ref{scale}] and [\ref{average}] and solving, we get
\begin{equation}
\alpha = \frac{1 + 2 \sigma}{1 + q}, \quad \beta = \frac{1 - \sigma}{1 + q}. 
\end{equation}

The Kasner solutions are summarized in Table~[\ref{tabl:kasner}]. From the exact solutions
we observe that in the case of cosmic strings, universe accelerates in one
direction and decelerates in the other two directions. Then, in the case of domain walls,
it decelerates in one direction and  accelerates in the other two directions. For LVMF,
it contracts in one direction and decelerates in the other two directions, and for
magnetic fields, universe expands along only one dimension, while it's static in the
other two dimensions.
\begin{table}
 \centering
 \begin{tabular}{|l|l|l|l|l|r|}
  \hline
  Matter            & $\sigma_0$ & $\alpha$ & $\beta$ \\
  \hline
  Cosmic String     &    1/2   & 4/3      &    1/3 \\
  Domain Walls      &   -2/5   & 2/9      &   14/9 \\
  LVMF              &   -1     &-1/3      &    2/3 \\
  Magnetic Field    &    1     & 1        &    0  \\
  \hline
 \end{tabular}
 \caption{Kasner type solutions at the asymptotic stable fixed points of the state space
          variables.}
 \label{tabl:kasner}
\end{table}

\section{A more realistic scenario}\label{sec:realscenario}
In this section we extend our analysis of the previous section by adding
the usual dust like dark matter and cosmological constant as dark energy to our
anisotropic matter types, in order to make it a realistic scenario, relevant
for the current evolution of our universe. We do our
analysis in $\tau-$time which is useful to find stable fixed points.
However, the form of the geodesic equations will be unchanged.
Here we analyse the asymptotic evolution numerically.

\begin{figure}
 \centering
 \includegraphics[width=0.84\textwidth]{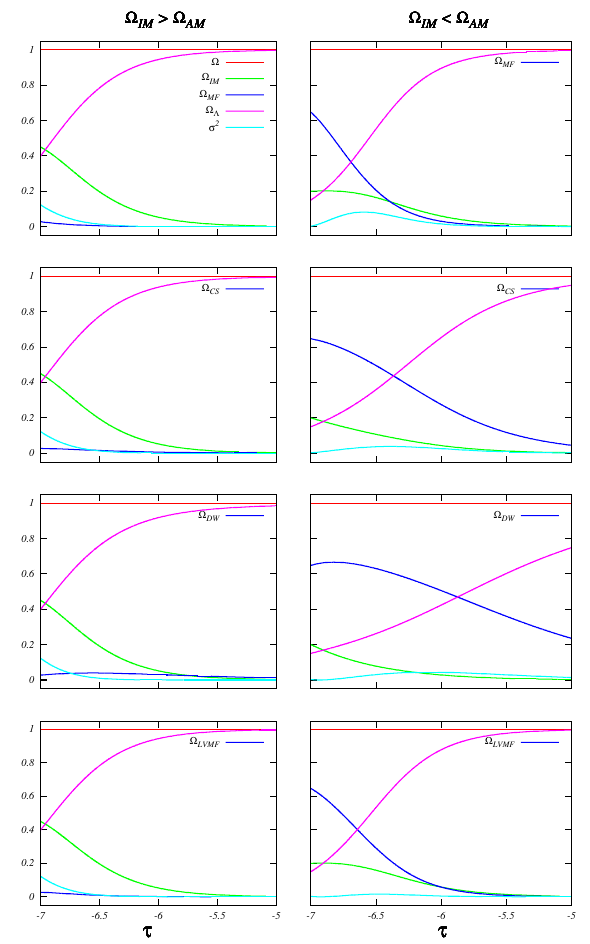}
 \caption{$\Omega$ vs $\tau$ : Late time behaviour of shear and density
          fractions of $\Lambda$CDM components in addition to
          the individual anisotropic sources studied in this work.
          Some representative initial values were used for the two
          cases of $\Omega_{IM} > \Omega_{AM}$ (\emph{left column}) and
          $\Omega_{IM} < \Omega_{AM}$ (\emph{right column}). The evolution
          shown here is from $\tau_{LSS} =-7$ to $\tau_0 = 0$ (today).}
 \label{fig:realmodel}
\end{figure}

The evolution equations in $\tau-$time for this model which includes anisotropic matter,
isotropic (dust like) dark matter and dark energy ($\Lambda$) are given by
\begin{eqnarray}
\frac{dH}{d\tau} &=& - (1+q) H \,, \nonumber \\
q &=& 2 \sigma^2 + \frac{1}{2} \left[\Omega_{IM} (1 + 3 w_{IM}) + \Omega_{AM} (1 + 3 w_{AM}) + \Omega_{\Lambda} (1 + 3 w_{\Lambda}) \right] \,, \nonumber \\
\frac{d\sigma}{d\tau} &=& (1+q) \sigma - 3 \sigma + (w_b - w_a) \Omega_{AM} \,,  \label{eq:EvolvEqnRealModel} \\
\frac{d\Omega_{IM}}{d\tau} &=& 2 (1+q) \Omega_{IM} - 3 (1 + w_{IM}) \Omega_{IM} \,, \nonumber \\
\frac{d\Omega_{\Lambda}}{d\tau} &=& 2 (1+q) \Omega_{\Lambda} \,, \nonumber \\
\Omega_{AM} &=& 1- (\Omega_{IM} + \Omega_{\Lambda} + \sigma^2) \,, \nonumber 
\end{eqnarray}
where,
\begin{eqnarray}
w_{IM} &=& 0 \,, \nonumber \\
w_{\Lambda} &=& - 1 \,, \nonumber\\
w_{AM} &=& (w_a + 2\,w_b)/3 \,, \nonumber \\
p_{IM} &=& w_{IM} \rho_{IM} \,, \nonumber \\
p_{\Lambda} &=& w_{\Lambda} \rho_{\Lambda} \,, \nonumber\\
p_{AM}^i &=& w_{AM}^i \rho_{AM} \,,\quad i = a,b \, \nonumber\\
\sigma &=& h/\sqrt{3} H \,, \nonumber\\
\Omega_i &=& \rho_i / 3 H^2 ,\quad i = IM, \Lambda, AM \,. \nonumber
\end{eqnarray}
The evolution of $\sigma, \Omega_{IM}, \Omega_{\Lambda}$ and $\Omega_{AM}$ are
given in Fig.~[\ref{fig:realmodel}] for the two cases
$\Omega_{IM} > \Omega_{AM}$  and $\Omega_{IM}< \Omega_{AM}$.
The late time behaviour in all these cases is almost same except for domain walls.
However small may be the value of $\Omega_{DW}$ to start with,
it turns out that $\Omega_{DW}$ tends to rise at intermediate times,
and can even dominates over isotropic matter at present time for some initial
values. This may not be a viable scenario as we know that the current dominant
matter content is of isotropic type. However the other anisotropic sources
can give rise to a viable scenario. For the actual evolution of an initial
ordinary (isotropic dark) matter dominated era with small fractions of anisotropic
matter and shear, to the current era of dark energy domination, the deceleration
parameter is found to evolve from 1/2 to -1 at late times, as expected.

\section{Temperature Patterns}\label{sec:cmbpattern}
For a Bianchi-I universe with anisotropic matter, along with the usual (isotropic)
dark matter and dark energy ($\Lambda$), we find the contribution to the CMB
temperature anisotropies due to these sources here.
It has been shown earlier that the total quadrupole anisotropy in the presence of a uniform
magnetic field, can be small compared to that obtained from
the standard $\Lambda$CDM model \cite{Campanelli0607}.
The temperature of the cosmic microwave background as a function of the
angular coordinates $\hat{n} = (\theta,\phi)$ on the celestial sphere
is given by \cite{Lim1999},
\begin{equation}
  T(\theta, \phi) = T_{LSS} \exp\left[ - \int_{\tau_{LSS}}^{\tau_0}(1+p) \, d\tau \right] \,
  \label{eq:Tdef}
\end{equation}
where $T_{LSS}$ is the mean isotropic temperature of the CMB at the surface
of last scattering, and
\begin{equation}
  p = \Sigma_{\alpha\beta}K^\alpha K^\beta \,.
 \label{eq:sdef}
\end{equation}
The direction cosines along a null geodesic $K^\alpha$ in
terms of the spherical polar angles ($\theta,\phi$) are given by
\begin{equation}
  \label{eq:Kdefs}
  K^\alpha(\tau_0) = \left( \cos\theta, \sin\theta \cos\phi , \sin\theta \sin\phi \right) \, .
\end{equation}
In our case of planar geometry, $\Sigma_{\alpha\beta}$ takes the form
\begin{equation}
  \Sigma_{\alpha\beta} = {\rm diag}( -2 \sigma, \sigma , \sigma ) \, .
  \label{eq:sigmadef}
\end{equation}  
Substituting Eq.~[\ref{eq:Kdefs}] and [\ref{eq:sigmadef}] into Eq.~[\ref{eq:sdef}] we get
\begin{equation}
  p = (1-3K_1^2)\sigma \ .
 \label{eq:pdef}
\end{equation}
The geodesic equation for $K_1$ is
\begin{equation}
\frac{d K_1}{d \tau} = 3(1-K_1^2)K_1\sigma\ , 
\label{eq:k1geoeq}
\end{equation}
Using Eq.~[\ref{eq:pdef}] and [\ref{eq:k1geoeq}] in Eq.~[\ref{eq:Tdef}], we get
\begin{equation}
T(\theta,\phi)= T_{LSS} \, \frac{e^{\tau_{LSS}-\tau_0} \, e^{2 \zeta_0}}{\sqrt{e^{6 \zeta_0} + (1 - e^{6 \zeta_0}) \cos^2\theta}},
\end{equation}
where $\tau_0$ ($=0$) and $\tau_{LSS}$ ($=-7$)
are the values of $\tau$ - today and at the time of decoupling, respectively - and
\begin{equation}
\zeta_0 = \int^{\tau_0}_{\tau_{LSS}} \sigma d\tau \,. 
\end{equation}

\begin{figure}
 \centering
 $
 \begin{array}{c c}
   \includegraphics[width=0.48\textwidth]{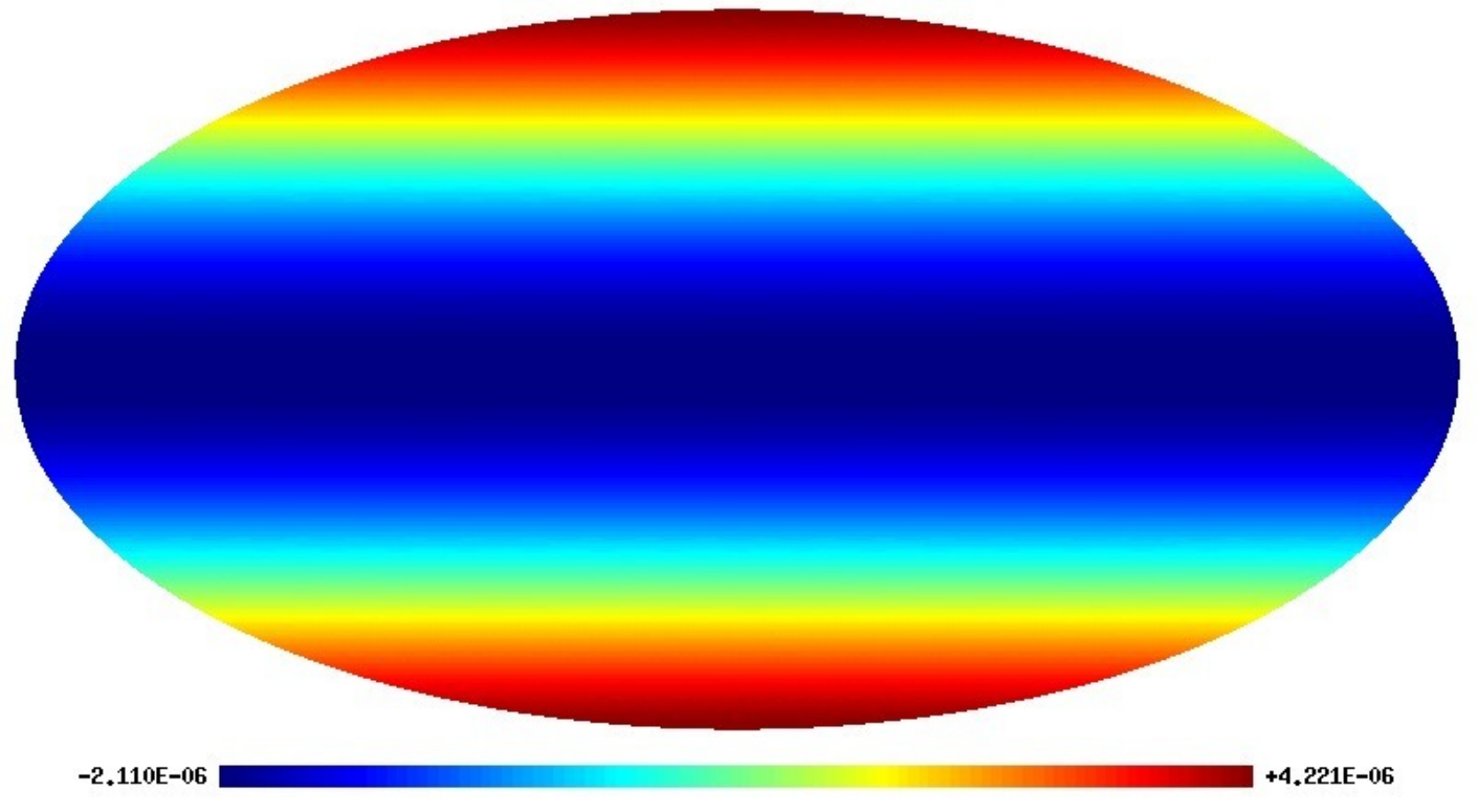} &
   \includegraphics[width=0.48\textwidth]{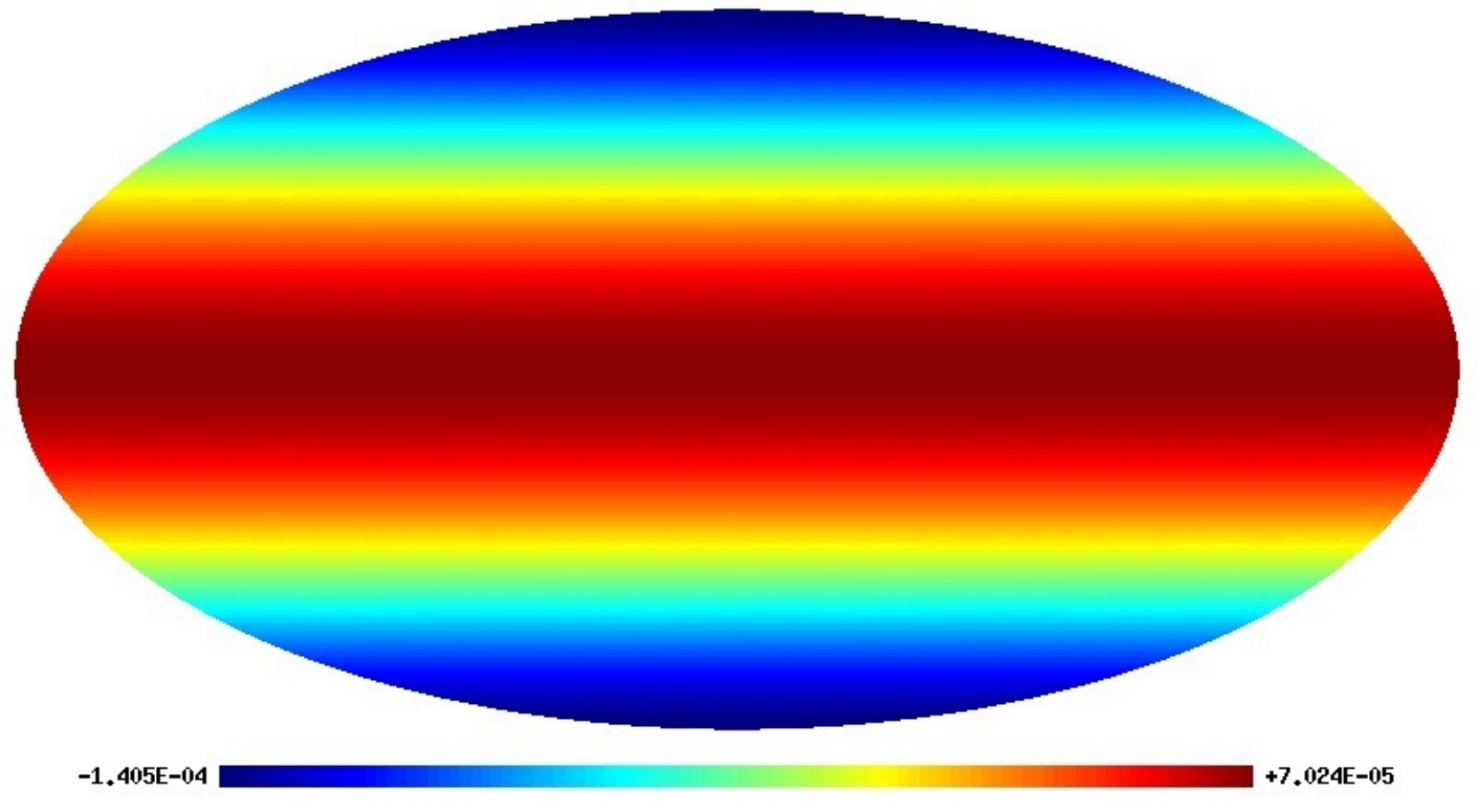} \\
   \includegraphics[width=0.48\textwidth]{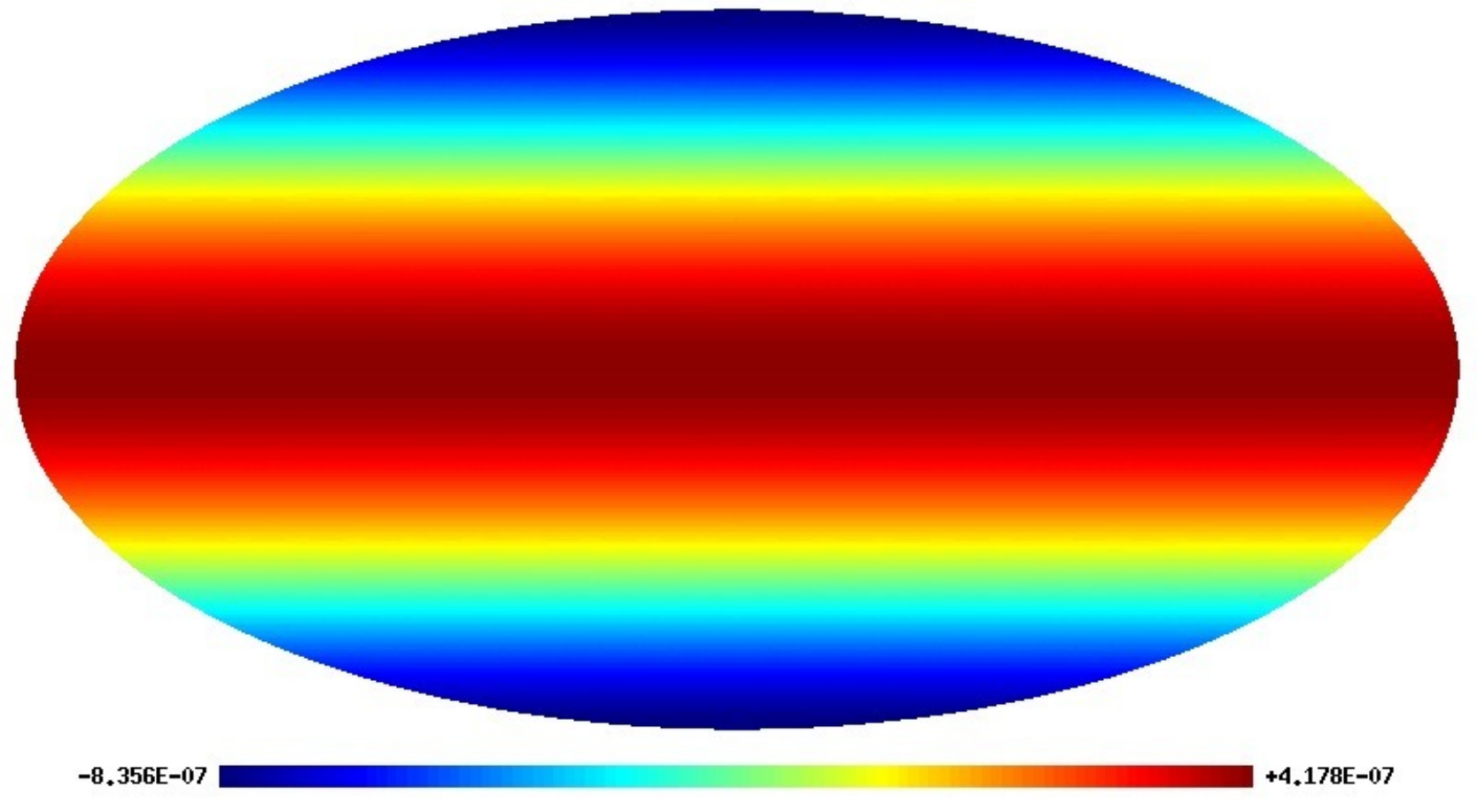} &
   \includegraphics[width=0.48
   \textwidth]{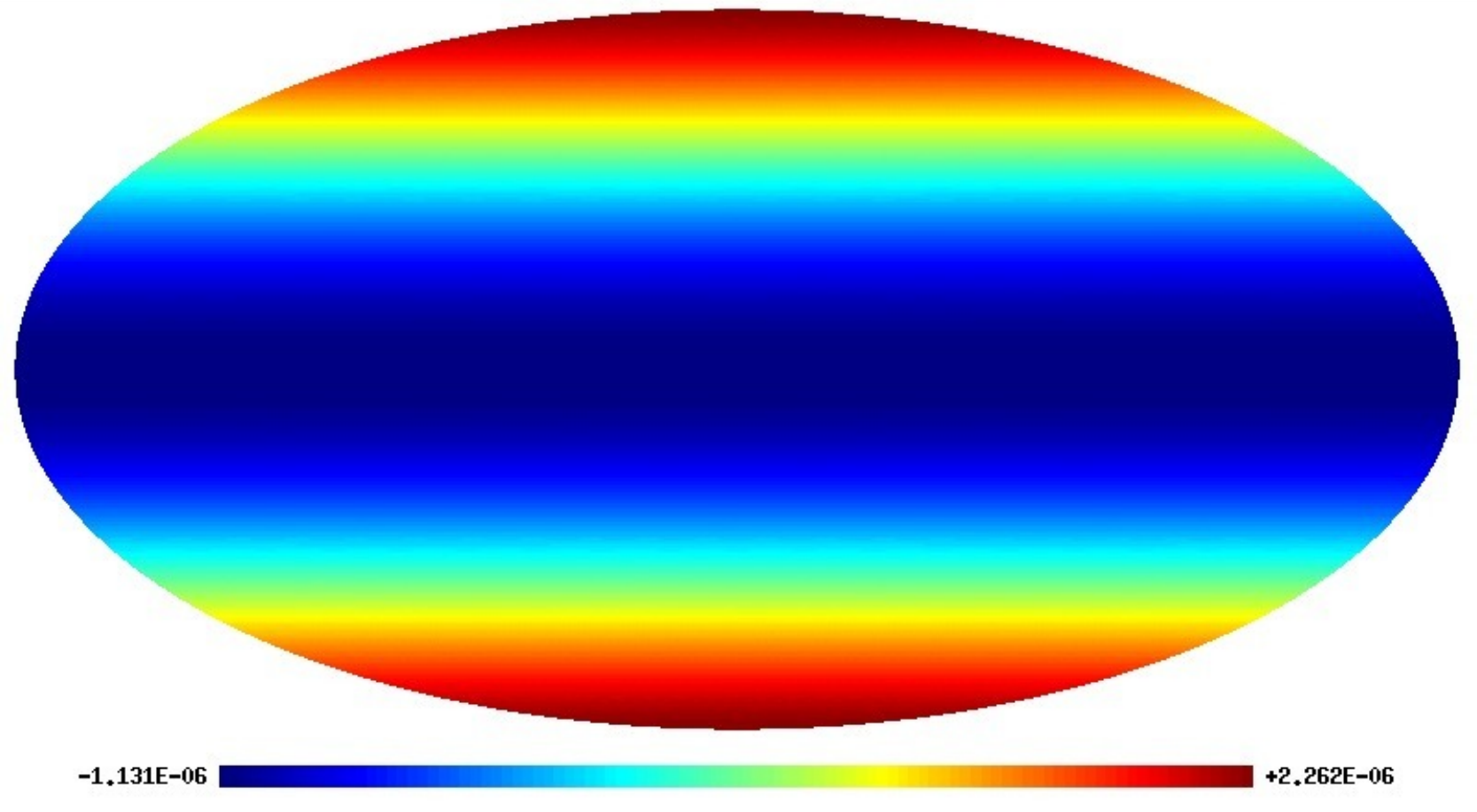}
 \end{array}
 $
  \caption{Temperature patterns ${T(\theta, \phi)}/{T_{LSS}}$ for
           Cosmic Strings (\emph{top left}), Domain walls (\emph{top right}),
           LVMF (\emph{bottom left}) and Magnetic field (\emph{bottom right}).
           Here we used $\sigma_{LSS}=10^{-8}$ for CS and DW, and
           $\sigma_{LSS}=10^{-5}$ for LVMF and MF.}
  \label{fig:map1}
\end{figure}

It turns out that the contribution is dominant to quadrupole for all the
four cases of anisotropic matter types considered here.
The CMB quadrupole temperature patterns
are shown in Fig.~[\ref{fig:map1}].  We find that all other higher
multipoles receive negligible contribution from these sources. Hence
we only show temperature patterns for $l=2$ here.
There could be a dipole anisotropy also in CMB due to spatially
inhomogeneous and anisotropy spaces such as, for example,
in Ref.~\cite{Alnes06}. But, since our matter types are comoving
sources and do not have peculiar motion, we do
not see a dipole contribution to CMB anisotropies due to these sources.
We also observe from  Fig.~[\ref{fig:map1}] that the temperature
maps for magnetic field and cosmic strings have similar patterns,
where as temperature maps for domain walls and LVMF show similar patterns.
This has to do with the signature of $\zeta_0$, which in turn is related to
$\sigma$ (see Fig.~[\ref{fig:sigma}]).

\section{SN Ia constraints on cosmic shear and anisotropy}\label{sec:snconstr}
In this section, we will test our anisotropic Bianchi-I model with anisotropic
sources using supernova data. We constrain the level of anisotropic matter density,
shear, and also determine the cosmic preferred axis, if present, in addition to the
current Hubble parameter, cold dark matter and dark energy (modeled as cosmological
constant) density fractions, using the Type Ia supernova (SN1a) objects in the
Union~2 compilation \cite{Union2ref}\footnote{The positions of the supernovae are in part
obtained from \url{http://vizier.u-strasbg.fr}, \url{http://www.cbat.eps.harvard.edu/lists/Supernovae.html}
and \url{http://sdssdp62.fnal.gov/sdsssn/snlist_confirmed_updated.php} web pages.}
For this section, we use the evolution equations as determined from the line element
\begin{equation}
ds^2 = dt^2 - a^2(dx^2 + dy^2) - b^2 dz^2
\end{equation}
which corresponds to the frame where the anisotropy axis is along the $z-$direction.
Thus, the mean scale factor and the eccentricity are given by
\begin{eqnarray}
A &=& (a^2 b)^{1/3} \nonumber \\
e^2 &=& 1 - \frac{b^2}{a^2} \, .
\end{eqnarray}

The redshift ($z$) and luminosity distance ($d_L$) of an SN1a object observed in the direction
$\hat{n} = (\theta,\phi)$ are given by \cite{AnisoDE}
\begin{equation}
1+z(\hat{n}) = \frac{1}{A} \frac{(1-e^2 \sin^2\alpha)^{1/2}}{(1-e^2)^{1/3}} \,,
\end{equation}
and
\begin{equation}
d_L(\hat{n}) = c\,(1+z) \int_{A(z)}^{1} \frac{dA}{A^2H} \frac{\left(1-e^2\right)^{1/6}}{\left(1-e^2\cos^2\alpha\right)^{1/2}}\,,
\label{eq:th-lum-dist}
\end{equation}
where $\alpha$ is the angle between the cosmic preferred axis ($\hat{\lambda}$) and
the supernova position ($\hat{n}$), and `$c$' is the speed of light. Hence,
$\cos\alpha = \hat{\lambda}\cdot\hat{n}$.
In order to obtain constraints on our Bianchi-I model with anisotropic matter and
shear in addition to $\Lambda$CDM components, we fit the measured distance modulus
of various SN1a objects that are provided in the Union~2 compilation to the
theoretical distance modulus, by minimizing the $\chi^2$ defined as
\begin{equation}
\chi^2 = \sum_{i} \frac{\left(\mu_i^d - \mu_i^{th} \right)^2}{\left(\delta\mu_i^d\right)^2} \, ,
\end{equation}
where $\mu_i^d$ is the measured distance modulus of an SN1a object from data,
$\mu_i^{th}$ is the theoretical distance modulus function involving various
cosmological parameters, $\delta\mu_i^d$ is the measured uncertainty in
the distance modulus of an SN1a object provided in the data, and the summation is over
all SN1a objects of the data (total 557 supernovae).

The theoretical distance modulus is defined as
\begin{equation}
\mu^{th} = 5\log\left(\frac{d_L}{10pc}\right)\,,
\label{eq:meas-dist-mod}
\end{equation}
where $pc$ in the denominator stands for `parsec', and $d_L$ is the luminosity
distance defined in Eq.~[\ref{eq:th-lum-dist}]. 
By using the parametrization $H = 100\,h\,Km.sec^{-1}/Mpc = k\,h$, where $k$
is dimensionfull and $h$ being the dimensionless Hubble
parameter, Eq.~[\ref{eq:meas-dist-mod}] can be further simplified to give
\begin{equation}
\mu^{th} = 5\log \tilde{d}_L + 5\log\left(\frac{c/k}{10^{-5}Mpc}\right) = 5\log \tilde{d}_L + \mu_0\,,
\end{equation}
where $\mu_0 =42.384$, and $\tilde{d}_L$, containing dimensionless quantities,
is given by
\begin{equation}
\tilde{d}_L (\hat{n}) = (1+z) \int_{A(z)}^{1} \frac{dA}{A^2\,h} \frac{\left(1-e^2\right)^{1/6}}{\left(1-e^2\cos^2\alpha\right)^{1/2}} \,.
\end{equation}
In this section, we use `$h$' to denote the dimensionless Hubble parameter. The
luminosity distance, $\tilde{d}_L$, (and thus the distance modulus $\mu$)
depends on all the cosmological parameters
$\{h_0(\equiv H_0), \sigma_0, \Omega_0^{AM}, \Omega_0^{IM}, \Omega_0^{\Lambda}, \hat{\lambda}=(l_A,b_A)\}$.
The minimization is done in conjugation with solving the evolution equations
\begin{eqnarray}
\frac{dA}{dz} &=& \frac{-f_1}{(1+z)(1+z-f_1\,f_2\frac{\sigma}{A})} \\
f_1 &=& \frac{(1-e^2\sin^2\alpha)^{1/2}}{(1-e^2)^{1/3}} \nonumber \\
f_2 &=& \frac{2+(e^2-3)\sin^2\alpha}{1-e^2\sin^2\alpha} \nonumber \\
(e^2)' &=& \frac{6\sigma}{A}(1-e^2) \nonumber \\
\frac{h'}{h} &=& -\frac{3}{2}\left( 1 + \sigma^2 + \bar{w}\,\Omega_{AM} -\Omega_{\Lambda} \right)  \nonumber\\
\sigma' &=& -\frac{3}{2}\left( \sigma(1-\sigma^2) -(\frac{2}{3}\delta_w+\bar{w}\,\sigma)\Omega_{AM} + \sigma\Omega_{\Lambda} \right)  \nonumber\\
\Omega_{AM}' &=& -3\,\Omega_{AM} \left( \bar{w} + \frac{2}{3} \delta_w\sigma -\bar{w}\,\Omega_{AM} + \Omega_{\Lambda} -\sigma^2\right) \nonumber \\
\Omega_{IM}' &=& 3\,\Omega_{IM} \left( \bar{w} \, \Omega_{AM} - \Omega_{\Lambda} + \sigma^2 \right) \nonumber \\
\Omega_{\Lambda}' &=& 3\,\Omega_{\Lambda} \left( 1 + \bar{w}\,\Omega_{AM} - \Omega_{\Lambda} +\sigma^2 \right) \nonumber
\end{eqnarray}
where $\bar{w} = (2\,w_a+w_b)/3$, $\delta_w = w_a - w_b$, $\sigma$ is the shear
and $\Omega_{AM}, \Omega_{IM}, {\Omega_\Lambda}$ are the fractional energy densities due
to anisotropic matter ($AM$), ordinary isotropic dark matter ($IM$) and the dark energy
($\Lambda$). The $'$  denotes a derivative with respect to `$\tau$'. The mean scale
factor `$A$' and $\tau-$time are related by ${dA}/{d\tau} = A$.
These equations are evolved from $z=[0,z_i^{SN}]$ for each supernova $i$ with the initial
conditions at $z=0$ as $A=1$, $e=0$ corresponding to the choice $a_0 = 1$ and $b_0 = 1$,
and random guess values for the parameters
$\{ h_0, \sigma_0, \Omega_0^{AM}, \Omega_0^{IM}, \Omega_0^{\Lambda}, \hat{\lambda} \}$
to do the $\chi^2$ minimization.
The deceleration parameter is given by $q_0 = h'/h|_{z=0}$.
These anisotropic matter source models in a Bianchi-I universe are
compared with the standard concordance model using the luminosity
distance relation given by
\begin{equation}
\tilde{d}_L = (1+z)\int_{0}^{z^{SN}} \frac{dz}{\sqrt{\Omega_0^{IM}(1+z)^3+\Omega_0^\Lambda}}\,,
\end{equation}
where $\Omega^{IM}+\Omega^\Lambda = 1$.

\begin{figure}
 \centering
 \includegraphics[width=0.9\textwidth]{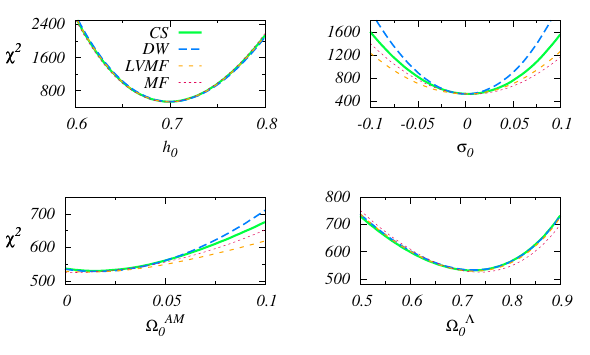}
 \caption{The behaviour of $\chi^2$ as a function of the parameters
          $ h_0, \sigma_0, \Omega_0^{AM}$ and $\Omega_0^{\Lambda}$
          for the anisotropic universe at it's minimum are shown here.
          $\Omega^{IM}_0$ dependence is not shown, treating it as a
          dependent parameter that can be estimated from the constraint
          equation $\Omega_{IM} = 1 - \sigma^2 - \Omega_{AM} - \Omega_\Lambda$.}
 \label{fig:sn-cosmopar-am}
\end{figure}

\begin{figure}
 \centering
 \includegraphics[width=0.9\textwidth]{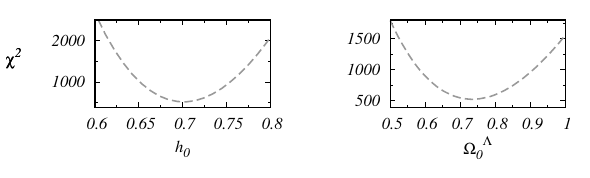}
 \caption{Same as Fig.~[\ref{fig:sn-cosmopar-am}], but for standard $\Lambda$CDM model
          parameters.}
 \label{fig:sn-cosmopar-lcdm}
\end{figure}

\begin{figure}
 \centering
 $
 \begin{array}{c c}
 \includegraphics[width=0.4\textwidth]{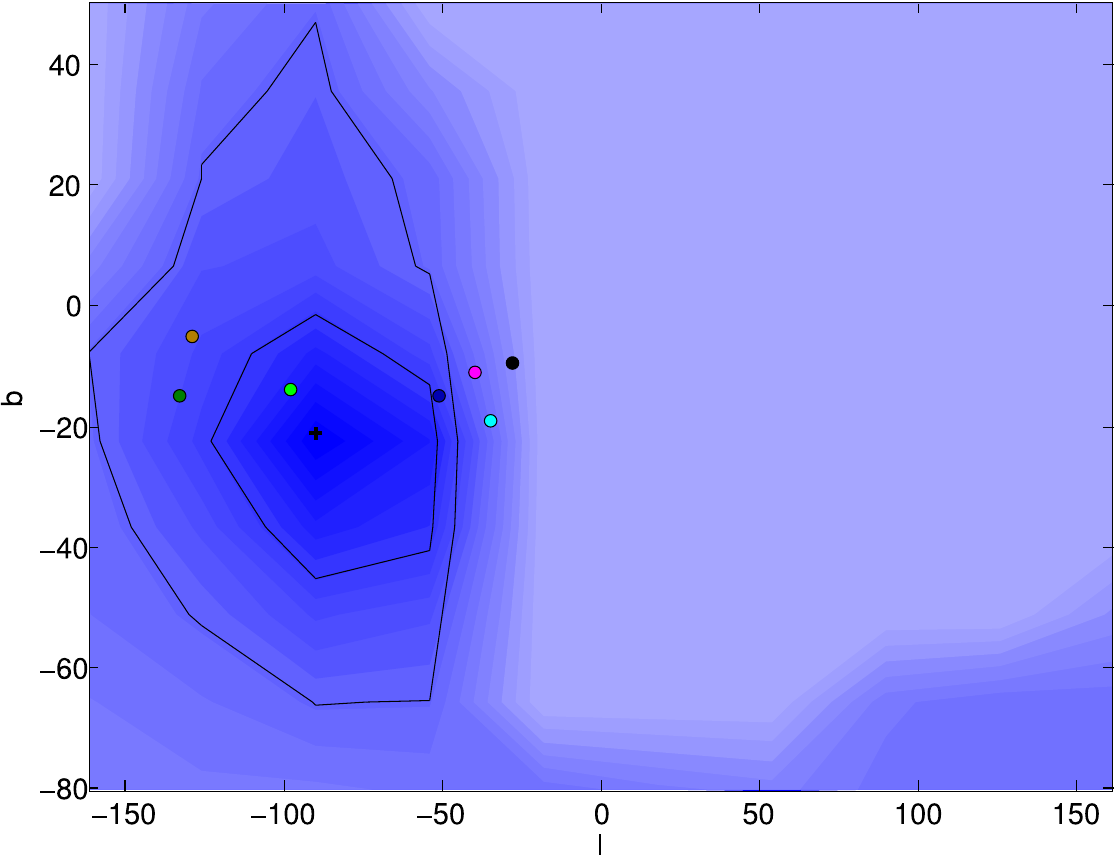} &
 \includegraphics[width=0.4\textwidth]{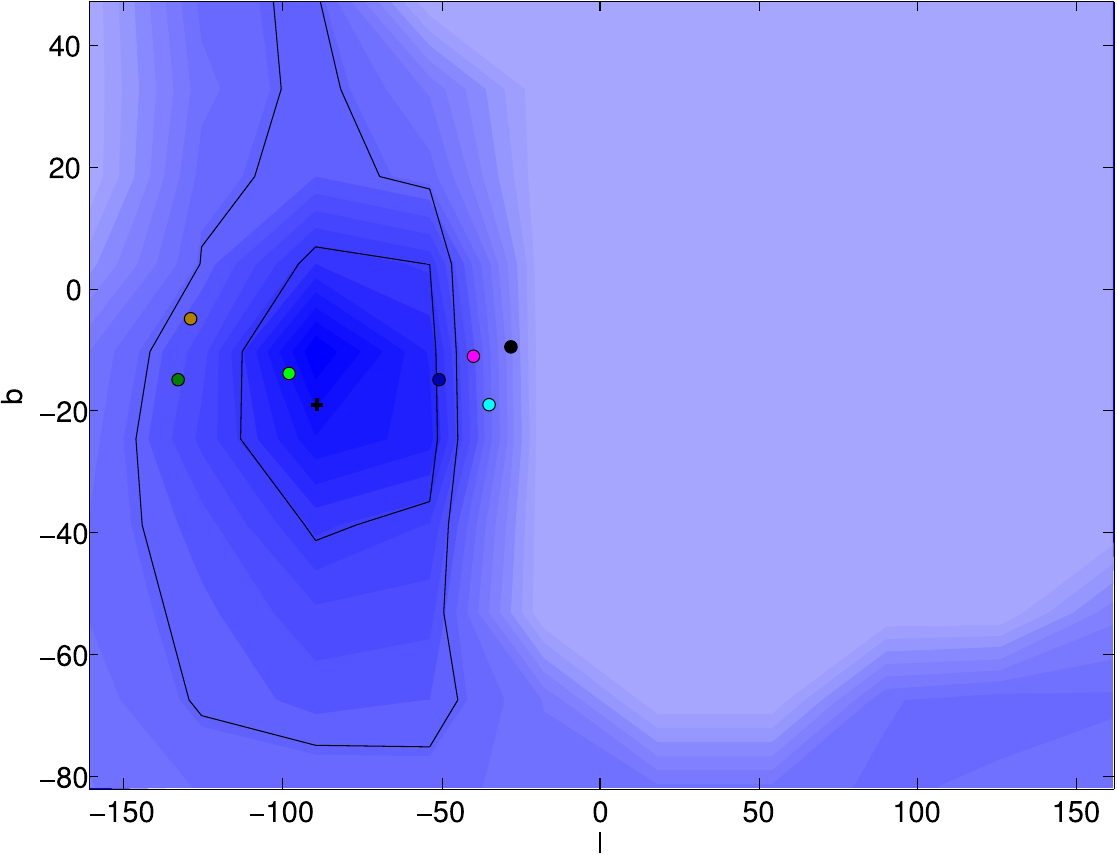} \\
 \includegraphics[width=0.4\textwidth]{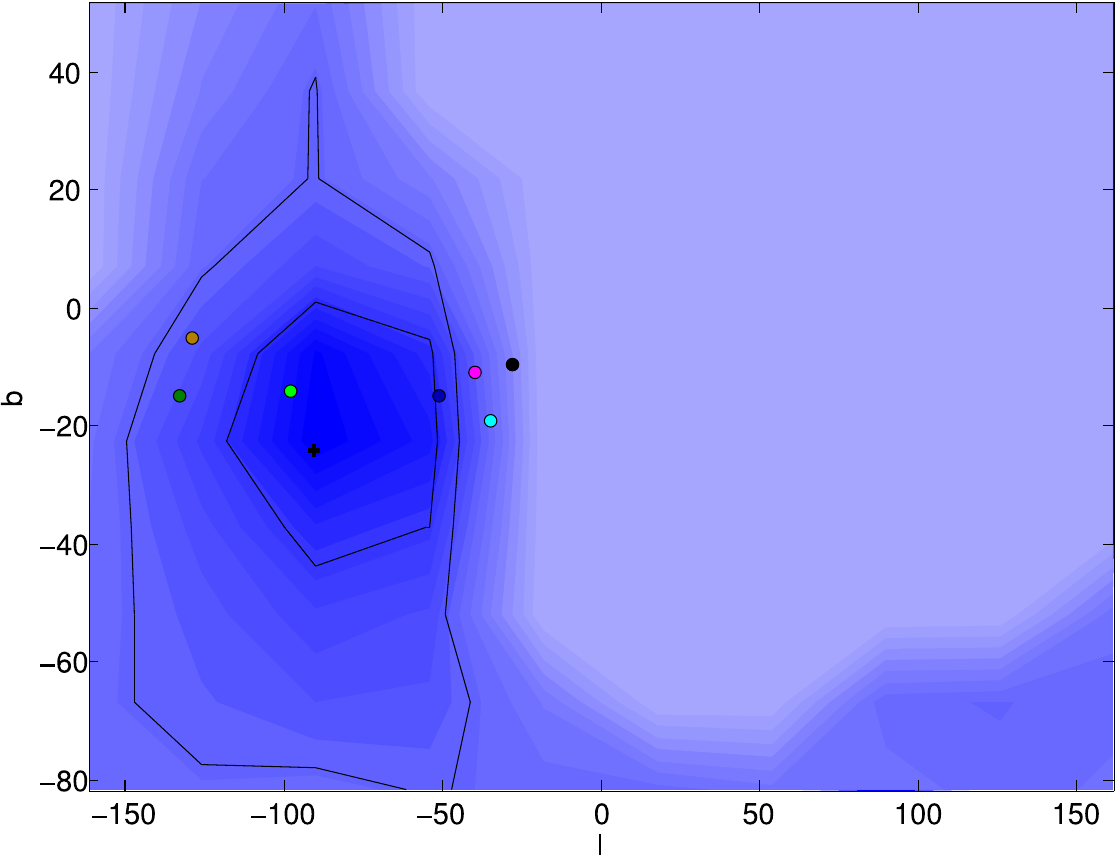} &
 \includegraphics[width=0.4\textwidth]{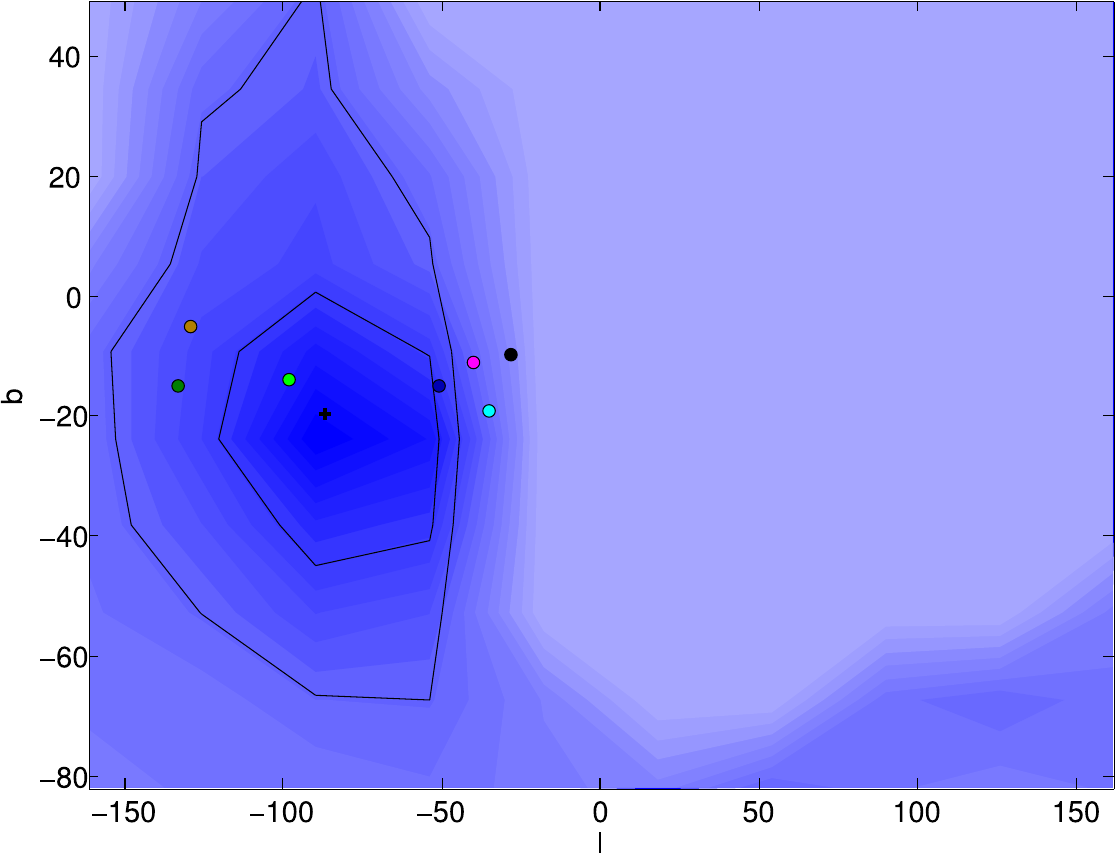}
 \end{array}
 $
 \medskip
 \includegraphics[width=0.78\textwidth]{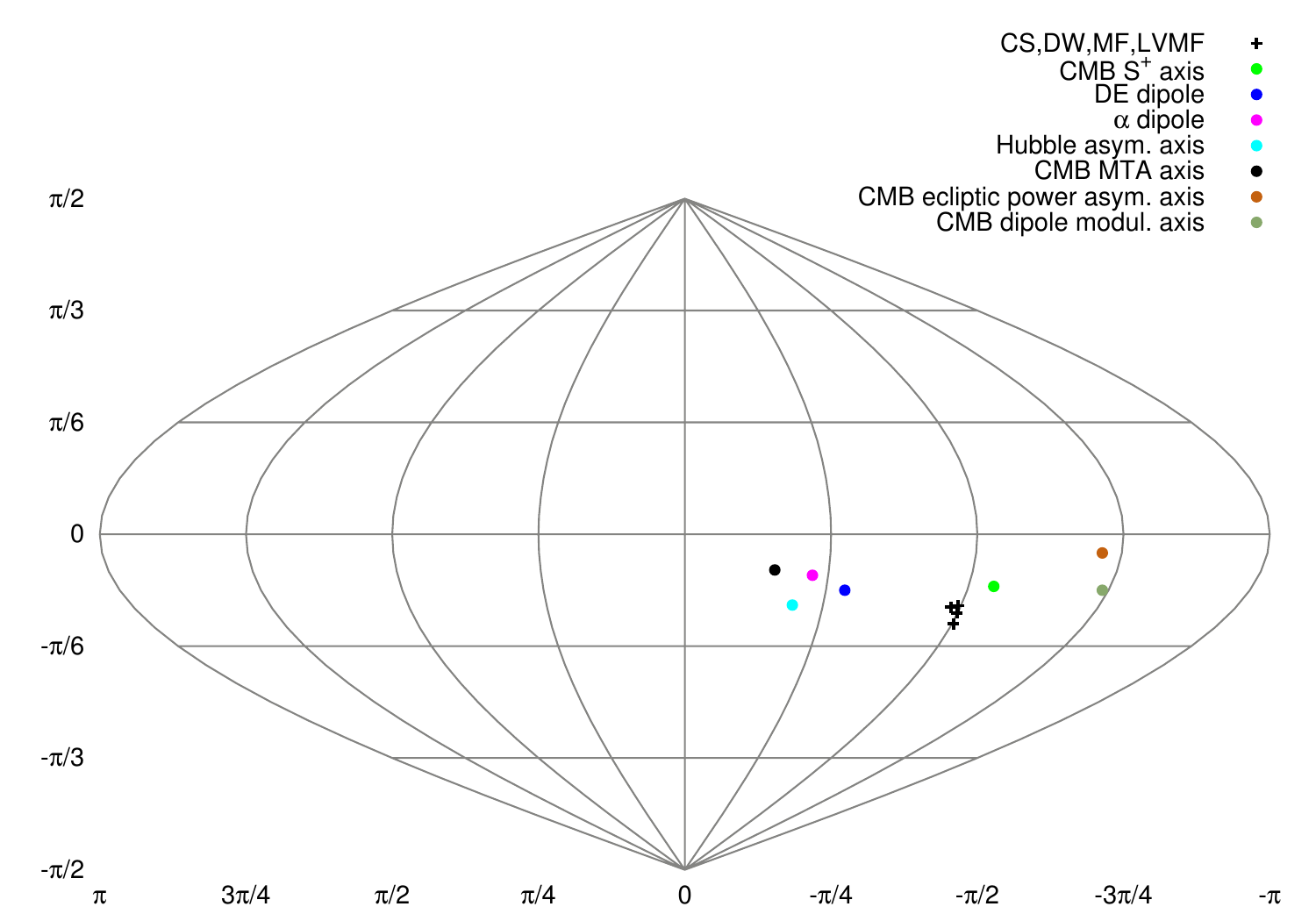}
 \caption{Plotted here are the anisotropic axes found in this work together with
          various anisotropic axes found in diverse cosmological observations.
          The reference co-ordinate system is the Galactic co-ordinate system $(l,b)$.
          The first four plots correspond to likelihoods of the anisotropic axis
          $\hat{\lambda}=(l_A,b_A)$ for cosmic strings, domain walls,
          magnetic fields and LVMF, respectively in the clockwise direction. The preferred
          axis specific to each of the anisotropic sources is plotted as a `$\bm{+}$', and
          coloured circles correspond to the other anisotropic axes.
          The $1\sigma$ and $2\sigma$ confidence regions are
          shown encompassed in black solid lines. The bottom figure shows, in perspective,
          all the axes on a sinusoidal projection of the celestial sky in galactic co-ordinates.
          These additional axes are tabulated in Table~[\ref{tab:aniso-axes}].}
 \label{fig:aniso-axes}
\end{figure}

The results from supernovae distance modulus fits to the models with
individual anisotropic sources in addition to cold dark matter and $\Lambda$,
and for the model with standard $\Lambda$CDM components alone, are given
in Table~[\ref{tab:sn-fit}].
The corresponding $\chi^2$ near it's minimum as a function of the cosmological
parameters in our anisotropic model, and only the $\Lambda$CDM parameters are
shown in Fig.~[\ref{fig:sn-cosmopar-am}], [\ref{fig:sn-cosmopar-lcdm}] and
[\ref{fig:aniso-axes}].
In those figures, note that the $\chi^2$ is shown only for the effective six parameters
$\{ h_0, \sigma_0, \Omega_0^{AM}, \Omega_0^{\Lambda}, \hat{\lambda}=(l_A,b_A)\}$
of our anisotropic model with individual anisotropic sources along with the standard model
components in a Bianchi-I background, and the two $\Lambda$CDM
parameters $\{ h_0, \Omega_0^{\Lambda}\}$ corresponding to the flat FRW
universe with the usual cold dark matter and dark energy, respectively. The
ordinary isotropic matter ($\Omega_{IM}$), is treated as dependent
quantity in both the cases that can be estimated from the corresponding
constraint equations.
On the whole, we find a marginal improvement of $\chi^2$ with our anisotropic
model compared to the standard flat $\Lambda$CDM model.

We find that the energy density fractions of the anisotropic sources'
considered in this work are consistent with zero with in a $1\sigma$ confidence level.
We also find that the anisotropic models considered here allow for
a very small, but non-zero, shear for our universe at present times.
So, a small non-zero energy density for these anisotropic sources, today,
may be plausible consistent with the error bars on them, in line with
the small non-zero shear today found in our anisotropic model fits to
SN1a data. The data also reveals a cosmic preferred axis for our universe,
independent of the anisotropic source model we used. This anisotropy axis
is also found to point in roughly the same direction as some
anisotropy axes found in other cosmological data that are tabulated in
Table~[\ref{tab:aniso-axes}].

As is evident from Fig.~[\ref{fig:aniso-axes}], there is a weak
dependence of $\chi^2$ on the anisotropy axis parameters $\hat{\lambda}=(l_A,b_A)$.
But we find the same preferred axis for the various equation of state
parameterizations corresponding to different anisotropic sources.
This might be indicative of a hidden preferred axis becoming explicit,
independent of the specific anisotropic parameterization used.
Both the weak dependence of anisotropy axis parameters on $\chi^2$,
and the overall improvement in  $\chi^2$, may be remedied by a future SN1a
data compilation which is homogeneous in both redshift and position spaces.

Interestingly the anisotropy axes we found here, are also approximately close to
other preferred directions found in diverse cosmological data \cite{AnisoAxes}.
Some (three) of the other anisotropy axes seem to be lying (just) outside of
the $2\sigma$ confidence level of our anisotropy axes plotted there.
But they all may agree with each other in direction within that limit,
as the $1\sigma$ bounds on these axes would overlap with our confidence contours
(see Ref.~\cite{PlanckAniso} and \cite{AnisoAxes} for $1\sigma$ bounds
on each the anisotropy axes of Table~[\ref{tab:aniso-axes}]).
The bottom plot of Fig.~[\ref{fig:aniso-axes}] shows all these diverse axes
in perspective. The CMB mirror parity symmetry axis \cite{PlanckAniso}
is the closest one to the axes we found here, at a mere angular separation of $10^\circ$.
All these axes pointing nearly in the same direction might be indicative of a
cosmic preferred axis for our universe.

\begin{table}
\centering
\begin{tabular}{c c c c c c}
\hline
                     & CS              & DW              & LVMF           &  MF             & $\Lambda$CDM \\
\hline
$h_0$                & 0.6971(31)      & 0.6980(30)      & 0.6982(30)      & 0.6977(31)      & 0.701 \\
$\sigma_0$           & 0.0088(44)      & 0.0068(47)      & 0.0060(40)      & 0.0070(44)      &   -    \\
$\Omega_0^{AM}$      & 0.010(17)       & 0.003(33)       & 0.000(11)       & 0.0013(52)      &   -    \\
$\Omega_0^{\Lambda}$ & 0.730(19)       & 0.735(22)       & 0.733(24)       & 0.735(16)       & 0.734 \\
$\Omega_0^{IM}$      & 0.260(36)       & 0.262(55)       & 0.267(35)       & 0.264(21)       & 0.266 \\
$\hat{\lambda}$      & (270.21,-21.09) & (270.92,-19.12) & (269.43,-24.03) & (273.11,-19.53) &   -    \\
$q_0$                & -0.600(37)      & -0.605(66)      & -0.599(41)      & -0.602(27)      & -0.601 \\
$\chi^2$             & 522.64          & 522.71          & 522.72          & 522.71          & 525.45 \\
$\Delta\chi^2$       & 2.81            & 2.74            & 2.73            & 2.74            &   -    \\
\hline
\end{tabular}
\caption{Supernovae constraints on the cosmological parameters of the two
         models, one with an anisotropic matter source in addition to $\Lambda$CDM
         components in a Bianchi-I universe and, the other with only $\Lambda$CDM
         components in a flat FRW background, respectively. The improvement $\Delta\chi^2$
         is with respect to $\Lambda$CDM model fitting (\emph{last column}).}
\label{tab:sn-fit}
\end{table}
\begin{table}[t]
\centering
\begin{tabular}{l l}
\hline
 Axis                      & ($l_A$,$b_A$) \\
\hline
Dark Energy dipole         & (309.4,-15.1)  \\
Fine structure constant    & (320.5,-11.7)  \\
dipole                     & \\
Asymmetric Hubble          & (325,-19)  \\
expansion axis             & \\
CMB Maximum Temperature    & (331.9,-9.6)  \\
Asymmetry axis             & \\
CMB ecliptic dipole power  & (231,-5)  \\
asymmetry axis             &  \\
CMB Dipolar modulation     & (218.9,-21.4)  \\
axis                       &  \\
CMB Mirror symmetry        & (262,-14)  \\
axis                       & \\
\hline
\end{tabular}
\caption{Preferred directions as observed in various cosmological data plotted in Fig.~[\ref{fig:aniso-axes}] in addition to the anisotropy axes found in our anisotropic model (see Table~[\ref{tab:sn-fit}]) are listed here \cite{PlanckAniso}, \cite{AnisoAxes}. The axis are given in Galactic co-ordinate system.}
\label{tab:aniso-axes}
\end{table}

\section{Conclusions}\label{sec:conclusion}
In this paper, we studied the evolution of Bianchi-I universe containing
different types of anisotropic matter sources. First we analyzed the evolution
of the full set of state space variables $H', h', \rho', k_0, k_1, k_2=k_3$ in cosmic
time. Depending on the initial conditions we found that there can be a sign change in
the shear. This may have some interesting implications
in early universe cosmology. Then we examined the evolution of dimensionless
variables viz., $\sigma$, $K_1$ and $K_3$ in dimensionless time $\tau$
using dynamical systems approach. In
the $\tau-$frame, we determined the fixed points of all the evolution
equations including the null geodesic equations. We also checked the stability of
the fixed points numerically.
We then investigated a more realistic scenario where we included
isotropic dust like (dark) matter and cosmological constant as
dark energy.
We found that the universe asymptotically evolves to a de~Sitter universe.
Then, in terms of $\tau-$time, we get an
analytic expression for the temperature anisotropies of the CMBR.
We generated the temperature patterns of the CMB for the four
anisotropic matter sources studied here. We found that the contribution to
the CMB temperature signal is mainly through the quadrupole. We also pointed out
the differences between the CMB temperature maps due to these anisotropic
matter types.

We then constrained the parameters of our anisotropic model using Union~2
type Ia supernova data.
We found that a very small, but non-zero shear could be present today
in our universe through our SN1a data constraints.
A cosmic preferred axis is also found from the data for all the
individual anisotropic matter source models in addition to $\Lambda$CDM
components in a Bianchi-I universe. The anisotropy axes we obtained
are almost same, independent of the anisotropic source we used.
It turns out that this axis is very close to the mirror symmetry
axis found in CMB data from Planck probe.
Interestingly enough, the axes we found here coincides with
the dipole axis found in a supernovae distance modulus - redshift fit
to an inhomogeneous universe that does not require dark energy to explain
the apparent late time acceleration of our universe \cite{Alens07}.
So, there seems to be some evidence that we live in an anisotropic
universe.

It will be interesting to generalize our results to other Bianchi
classes - A and B. In case of other Bianchi models the state space
would be larger and we need to find different subspace of the state space
and their stable fixed points. It will also be interesting to study the
CMB polarization anisotropy in these models.

\subsection*{Acknowledgements}
S.T. would like to acknowledge Department of Science and Technology, India
for financial support. We thank Santanu Das for sharing his MCMC cosmological
parameter estimation code, and acknowledge the use of \texttt{CosmoloGUI}\footnote{\url{http://www.sarahbridle.net/cosmologui/}},
a plotting interface to the popular \texttt{CosmoMC} output chains.
We also acknowledge the use of \texttt{HEALPix}\footnote{\url{http://healpix.jpl.nasa.gov/}} \cite{Healpix}
and Eran Ofek's MATLAB routine \texttt{coco.m}\footnote{\url{http://www.weizmann.ac.il/home/eofek/matlab/}}
for astronomical co-ordinate conversion of SN1a positions, in this work.
We thank the anonymous referee for his comments which greatly
helped in improving the clarity and presentation of this work.

\appendix
\appendixpage
\addappheadtotoc

\section{Anisotropic sources}
\label{apdx:A}
A network of domain walls can be formed during the phase transition 
in the early universe by spontaneous breaking of a $Z_2$ symmetry,
separated by a distance of the order of correlation length \cite{CosmoDefect}.
In a self interacting real scalar field theory for a domain wall
(in $yz-$plane), the energy momentum (e-m) tensor takes the form
\begin{equation}
T^{\mu(DW)}_{\nu}=A(x) \, \textnormal{diag}(1,0,1,1),
\end{equation} 
where A(x) has a bell shaped distribution around x=0.  The wall can be made 
thin by appropriately tuning the coupling strength and the vacuum expectation value 
of the self interacting scalar field.    

To find the form of the e-m tensor for a network of domain walls, let us
consider a collection of $N$ planar domain walls (with walls in the $yz-$plane)
within a box of volume $V$. Here we take the side of the box
to be much larger than the correlation length of the vacuum expectation value
of the scalar field. For this configuration e-m tensor depends on `$x$' alone.
Let us assume that the walls of this stack reside at points $x^i (i=1,2,\dots,N)$ on
the $x-$axis. Then the total e-m tensor for such a network of non-interacting walls is
given by
\begin{equation}
T_{\nu}^{\mu}= \sum_{i=1}^N T^{\mu(DW)}_{\nu}(x^i-x). 
\end{equation} 

In the case of a large N, we can have a function g(x) which is the average
number of walls per unit length in the range $x$ and $x+dx$. It is normalized
to satisfy $\int g(x) dx = N$. Then the average e-m tensor of this configuration
is given by
\begin{equation}
<T_{\nu}^{\mu}>= \frac {\int dx \int dx' g(x') T^{(DM)\mu}_{\nu}(x-x')}{\int g(x) dx}.
\end{equation} 
With an average distance between the walls as `$d$', we can approximate the
average e-m tensor as
\begin{equation}
<T_{\nu}^{\mu}> \simeq \frac{1}{d} D_{\nu}^{\mu},
\label{emtensor}
\end{equation} 
where $D_{\nu}^{\mu}= diag (\eta, 0, \eta, \eta)$ and
$\eta=\int A(x)\,dx$ is the surface energy density.

Now we can use appropriate Lorentz transformation to get the e-m tensor for
the case when domain walls are moving in the $x-$direction with an average velocity
$\beta$.  In this case $D_{\nu}^{\mu}$ becomes
\begin{eqnarray}
 D_{\nu}^{\mu} = \left(
                 \begin{array}{cccc}
                   \gamma^2 \eta & -\gamma^2 \beta \eta & 0 & 0 \\
                   -\gamma^2 \beta \eta & \gamma^2 \beta^2 \eta & 0 & 0 \\
                   0 & 0 & \eta & 0 \\
                   0 & 0 & 0 & \eta
                 \end{array}
                 \right)
\end{eqnarray}
where $\gamma = \sqrt{1-\beta^2}.$ For slowly moving domain walls, e-m tensor can be approximated by  
\begin{equation}
<T_{\nu}^{\mu}> \simeq \frac{1}{d} \, \textnormal{diag}(\eta, 0, \eta, \eta).
\end{equation}    
Such a network of slowly moving domain walls may be present now, for example,
due to an interaction of dark matter with the domain walls \cite{Massarotti1991}.
Distortion of the cosmic microwave background is negligible in that toy model.
This network of stacked non-relativistic domain walls can be described as a
perfect fluid with equation of state $p = (p_a+2p_b)/3 = -\frac{2}{3d}\eta$.

Similarly, it is well known that a network of cosmic strings can be formed
during the phase transition in the early universe when a U(1) symmetry is
broken. The e-m tensor due to an infinite string with mass per unit length
$\mu$ along the $x-$direction is given by \cite{CosmoDefect}
\begin{equation}
T^{\mu(CS)}_{\nu}= \mu \, \delta(y) \, \delta(z) \, \textnormal{diag}(1,1,0,0) \,.
\end{equation} 
Analogous to our construction of domain wall network one can easily construct
the average e-m tensor for a network of slowly moving cosmic strings along a
particular direction which can be approximated by
\begin{equation}
T^{\mu}_{\nu} \simeq \frac{1}{d^2} diag(\mu,\mu,0,0),
\end{equation} 
where $d$ is the average separation between the cosmic strings. Such a network of
non-relativistic cosmic strings can be described as a perfect fluid with an equation
of state $p = (p_a+2p_b)/3 = -\frac{1}{3d^2}\mu$.

The actual evolution of cosmic string and domain wall networks are complicated.
In this paper, we do not consider the nonlinear evolution of these networks with
effects such as formation of intersection and loops, and any gravitational effects
such as collapse and emission of gravitational radiation.

\section{Stability analysis}
\label{apdx:B}
Let us denote $\sigma_0, K_1^0$ and $K_2^0$ to be the fixed points
of Eq.~[\ref{eq:sigma}], [\ref{eq:omega}] and [\ref{eq:momentum}].
Let us consider perturbations around these fixed points as
$\sigma_0+\Delta \sigma$, $K_1^0+\Delta K_1$ and $K_2^0+\Delta K_2$.
Keeping terms only upto first order in perturbation, we get
\begin{eqnarray}{\scriptsize
 \frac{d}{d \tau} \left[
   \begin{array}{c}
     \Delta \sigma \\
     \Delta K_1\\
     \Delta K_2
   \end{array}
 \right] =
 \left[
   \begin{array}{ccc}
     -\frac{3}{2}(1-w)(1-3 \sigma_0^2) + 2 (w_a - w_b) \sigma_0  &  0  &  0 \\
     3 (1-(K_1^0)^2) K_1^0  &  3 (1-3(K_1^0)^2) \sigma_0  &  0 \\
     -3 (K_1^0)^2 K_2^0  &  -6 K_1^0 K_2^0 \sigma_0  &  -3 (K_1^0)^2 \sigma_0
   \end{array}
 \right]
 \left[
   \begin{array}{c}
     \Delta \sigma \\
     \Delta K_1 \\
     \Delta K_2
   \end{array}
 \right]}\,.
 \label{eq:fp_anls}
\end{eqnarray}
The three eigenvalues of the above matrix equation viz.,
$\gamma_1, \gamma_2, \gamma_3$ are given by
\begin{eqnarray}
 \gamma_1 &=& -\frac{3}{2}(1-w)(1-3 \sigma_0^2) + 2 (w_a - w_b) \sigma_0 \nonumber \\
 \gamma_2 &=& 3 (1-3(K_1^0)^2) \sigma_0 \\
 \gamma_3 &=& -3 (K_1^0)^2 \sigma_0 \,.  \nonumber
 \label{eq:fp_eigen}
\end{eqnarray}
These eigenvalues determine whether a fixed point is stable or unstable.
For the shear parameter, $\sigma$, all possible fixed points are listed in
Table~[\ref{tab:fp_sigma}], and the complete list of fixed points of $K_i$,
as dictated by the eigenvalues $\gamma_1, \gamma_2$ and $\gamma_3$, are
given in Table~[\ref{tab:fp_moment}].
\begin{table}
 \centering
 \begin{tabular}{c c c c}
  \hline
    & $\sigma_0$ & $g$ \\
  \hline
  \multirow{3}{*}{CS} &  1  &  2   & U \\
                      & -1  &  6   & U \\
                      & 1/2 & -3/2 & S \\
  \hline
  \multirow{3}{*}{DW} &  1   & 7      & U \\
                      & -1   & 3      & U \\
                      & -2/5 & -21/10 & S \\
  \hline
  \multirow{3}{*}{LVMF} &  1 & 4 & U \\
                        & -1 & 0 & S \\
                        & -1 & 0 & S \\
  \hline
  \multirow{3}{*}{MF} &  1 & -2 & S \\
                      & -1 &  6 & U \\
                      &  2 &  3 & U \\
  \hline
 \end{tabular}
 \caption{The full list of fixed points for the shear parameter, $\sigma$,
          are shown here, pointing out the stable (S) and unstable (U) values.
          Here `$g$' corresponds to the constant coefficient in equation
          $\frac{d \Delta \sigma}{d \tau} = g \,\Delta \sigma$
          from Eq.~[\ref{eq:fp_anls}].}
 \label{tab:fp_sigma}
\end{table}

\begin{table}
 \centering
 \begin{tabular}{c c c c c c c c}
 \hline
    & $\sigma_0$ & $K_1^0$ & $K_2^0$ & $\gamma_1$ & $\gamma_2$ & $\gamma_3$ \\
 \hline
 \multirow{2}{*}{CS} & \multirow{2}{*}{1/2} & $\pm 1$ &       0         & -3/2 & -3   & -3/2 & S \\
                     &     &    0    & $\pm 1/\sqrt{2}$ & -3/2 &  3/2 &  0   & U \\
 \hline
 \multirow{2}{*}{DW} & \multirow{2}{*}{-2/5} & $\pm 1$ &       0         & -21/10 & 12/5 & 6/5  & U \\
                     &      &    0    & $\pm 1/\sqrt{2}$ & -21/10 & -6/5 &  0   & S \\
 \hline
 \multirow{2}{*}{LVMF} & \multirow{2}{*}{-1} & $\pm 1$ &       0         & 0 & 6  & 3 & U \\
                       &    &    0    & $\pm 1/\sqrt{2}$ & 0 & -3 & 0 & S \\
 \hline
 \multirow{2}{*}{MF} & \multirow{2}{*}{1} & $\pm 1$ &       0         & -2 & -6 & -3 & S \\
                     &   &    0    & $\pm 1/\sqrt{2}$ & -2 &  3 &  0 & U \\
 \hline 
 \end{tabular}
\caption{The full list of fixed points corresponding to the set of equations
         Eq.~[\ref{eq:fp_anls}] is shown here. In the last column, an
         `S' denotes a stable fixed point and `U' denotes an unstable fixed point.}
\label{tab:fp_moment}
\end{table}

\end{document}